\documentclass[12pt]{iopart}
\pdfoutput=1
\usepackage{iopams}
\usepackage{graphicx}
\usepackage{booktabs} 
\usepackage[latin1]{inputenc} 

\begin{document}

\title[Symmetry-breaking transitions...]{Symmetry-breaking transitions in networks of nonlinear circuit elements}

\author{Martin Heinrich$^{1,2}$}
\author{Thomas Dahms$^{1}$}
\author{Valentin Flunkert$^{1}$}
\author{Stephen W. Teitsworth$^{2}$}
\author{Eckehard Sch{\"o}ll$^{1}$}

\address{$^{1}$Institut f{\"u}r Theoretische Physik, Technische
  Universit{\"a}t Berlin, 10623 Berlin, Germany}
\address{$^{2}$Department of Physics, Duke University, Box 90305,
  Durham, North Carolina 27708-0305, USA}
\ead{schoell@physik.tu-berlin.de}

\begin{abstract}
  We investigate a nonlinear circuit consisting of $N$ tunnel diodes
  in series, which shows close similarities to a semiconductor
  superlattice or to a neural network.  Each tunnel diode is modeled
  by a three-variable FitzHugh-Nagumo-like system.  The tunnel diodes
  are coupled globally through a load resistor.  We find complex
  bifurcation scenarios with symmetry-breaking transitions that
  generate multiple fixed points off the synchronization manifold. We
  show that multiply degenerate zero-eigenvalue bifurcations occur, which
  lead to multistable current branches, and that these
  bifurcations are also degenerate with a Hopf bifurcation. These
  predicted scenarios of multiple branches and degenerate
  bifurcations are also found experimentally.
\end{abstract}

\pacs{05.45.-a, 02.30.Oz, 72.20.Ht, 73.50.Fq, 85.30.-z}

\submitto{\NJP}

\section{Introduction}

Nonlinear circuit elements, in particular semiconductor
nanostructures, are known to exhibit a wide range of complex
spatio-temporal patterns. Among these structures, the superlattice
\cite{ESA70,SCH98,SCH00,WAC02,DAN03,AMA04,BON05,SCH09} is a prominent
example. It consists of alternating layers of two semiconductor
materials with different band gaps. As a consequence this leads to an
energy band scheme with a periodic sequence of potential barriers and
quantum wells. The electrons are localized in the quantum wells if the
potential barriers are sufficiently thick. This gives rise to
sequential resonant tunneling of electrons between the individual
quantum wells. Applying a dc bias, various nonlinear charge transport
phenomena may occur, including negative differential conductance
(NDC). The nonlinearity may induce spatio-temporal patterns, examples
being stationary or traveling high and low field domains. Stationary
domains are manifested as multistable branches in the current-voltage
($I$-$V$) characteristics, where each branch corresponds to the
localization of the domain wall in a certain quantum well of the
superlattice. In this work we focus on current branches, which form
saw-tooth like patterns in the $I$-$V$ characteristic when the voltage
is swept up or down
\cite{ESA74,GRA91a,PRE94,KAS94,AMA01,HIZ06,XU07,XU09}.

\begin{figure}[ht]
  \includegraphics[width=\textwidth]{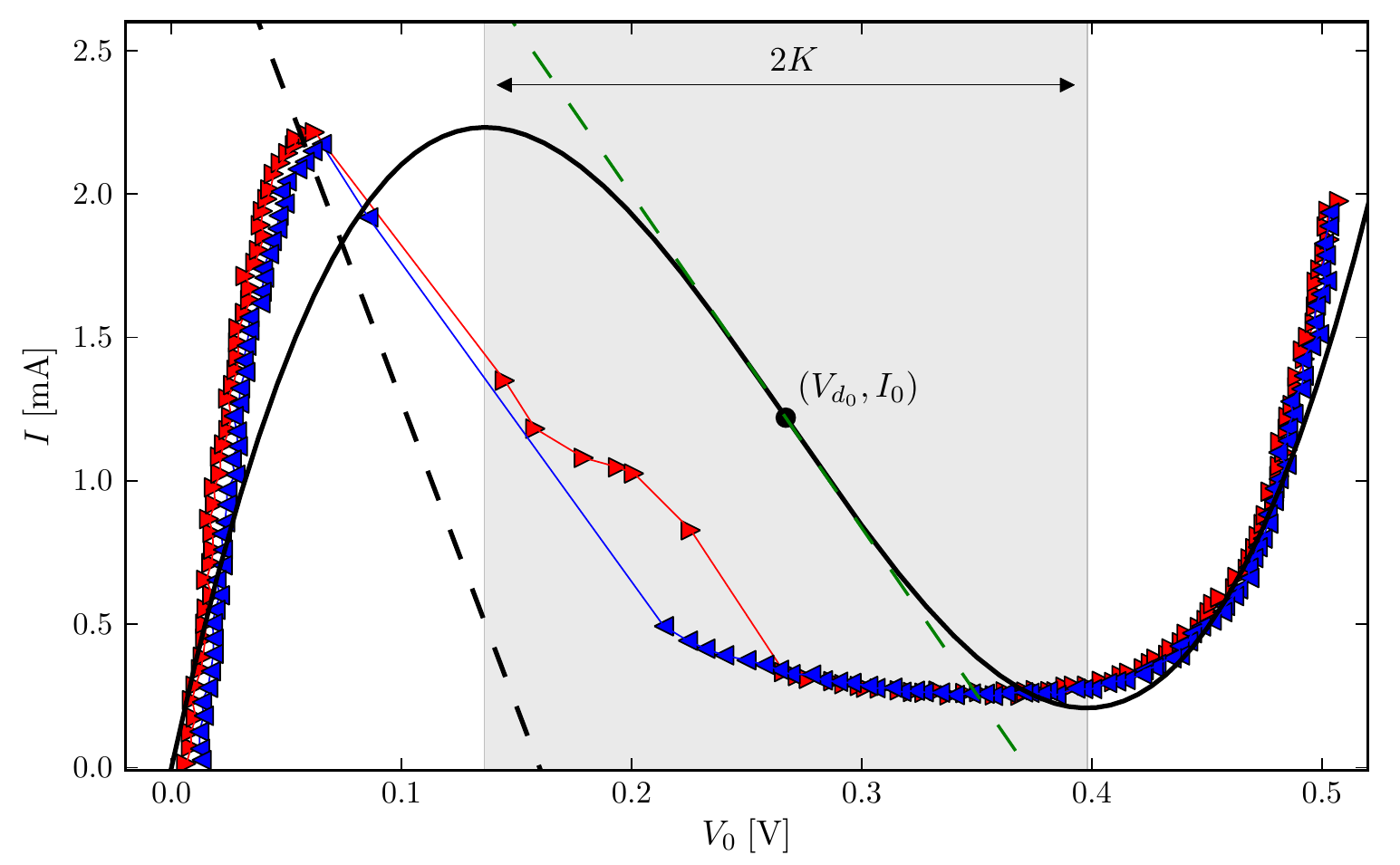}
  \caption{Current ($I$) -- voltage ($V$) characteristics of a tunnel
    diode. Measured data are denoted by red (voltage sweep-up) and
    blue (sweep-down) triangles. The regime of negative differential
    conductance (NDC) of the schematic $I$-$V$ curve (black) is marked
    by shading. The black and green dashed lines correspond to the
    load line at bias voltage $V_0 = 0.16$ V and load resistance
    $R=47$~$\Omega$, and the slope $-1/\rho$ at the inflection point
    $(V_{d_0}, I_0)$, respectively.}
  \label{img:bascufi}
\end{figure}

We aim to understand the phenomenon of current branches occurring in
superlattices and other semiconductor nanostructures by constructing a
breadboard equivalent to the basic electronic transport
characteristics in these devices. In fact, we will show that the
electronic transport in superlattices, which leads to current
branches, can be approximated by an equivalent circuit consisting of
tunnel diodes connected in series. Tunnel diodes are nonlinear circuit
elements, with similar $N$-shaped current-voltage characteristics as
superlattices and double barrier resonant tunneling diodes ($DBRT$)
\cite{KAZ71,TSU73,SCH00,SCH02,ROD03,STE07,MAJ09}, which can also show
a regime, where the current $I$ decreases with increasing voltage $V$.
This is depicted in Fig.~\ref{img:bascufi}, where the NDC regime of
the schematic $I$-$V$ curve is shaded for illustration. If the tunnel
diode is operated in a circuit with bias voltage $V_0$ and load
resistor $R$, the load line (dashed black) is given by $I=(V_0 - V)/R$
by Kirchhoff's laws.

In this work we investigate a chain of tunnel diodes
connected in series which results in a global (mean field) coupling. The
circuit equations, which we derive in Sec.~\ref{sec:model}, model the
NDC elements by a three-variable extension of the simplified
FitzHugh-Nagumo (FHN) system. The model reduces to the latter for a certain
choice of circuit parameters. Already in 1962, Nagumo et al. pointed out that
a model of a single neuron introduced by FitzHugh \cite{FIT61} can be
approximated by an electrical circuit containing a tunnel diode
\cite{NAG62}. Therefore, our results do not only have important
applications to electronic transport in NDC devices such as
superlattices, but also to the dynamics of neurosystems
\cite{IZH07,SCH07}.

In addition to the neuron and superlattice analogies, such NDC elements,
especially tunnel diodes, show promising applications in the design of
multi-junction solar cells \cite{HEN80a,GUT06,SZA08,HER08}, for
microwave and TeraHertz applications \cite{DES00a, DES01a, CID01,
  CID03, ASA08, MAG09}, in tunneling-based static random access memory
(TSRAM) \cite{WAG98,WAG99} and in digital logic circuits based on the
monostable-bistable transition logic element (MOBILE)
\cite{STO01,SUD04,APP08}.

The aim of this work is to study the bifurcations in a network of
tunnel diodes, both theoretically and experimentally. At the same time
this will offer insight into the dynamics of neural networks and
electronic transport in superlattices. We find Hopf bifurcations
degenerate with zero-eigenvalue bifurcations that are pitchfork
bifurcations for two coupled tunnel diodes, and pitchfork or
transcritical bifurcations for larger numbers of coupled tunnel
diodes. These degenerate bifurcations represent symmetry-breaking
nonequilibrium phase transitions \cite{SCH87}, where the symmetric,
synchronized state becomes unstable and two stable skew-symmetric
states arise. These symmetry-breaking bifurcations result in current
branches in the serial array of tunnel diodes: by increasing the bias
voltage, each tunnel diode subsequently passes through the NDC regime,
where oscillations arise.

Hopf-pitchfork bifurcations, which we find in two coupled tunnel diode
elements, were recently observed in a
spatio-temporal FHN-system \cite{MAU09} and in a pacemaker system
\cite{STI01a} and were investigated theoretically in
\cite{LAN98b,ALG00,ALG00a,KRA07}. The reason for the simultaneous
occurrence of Hopf and pitchfork bifurcations in our system of coupled
tunnel diodes is the coexistence of a reflection symmetry and
identical Hopf bifurcations in each of the single 3-variable tunnel
diode systems.

Pitchfork bifurcations usually arise in systems with $Z_2$-symmetry
and are generic in these systems.  Introducing a heterogeneity breaks
the symmetry and leads to an imperfect pitchfork bifurcation, i.e., in
systems without a $Z_2$ symmetry pitchfork bifurcations are of
codimension-two. The Hopf-pitchfork bifurcation can be unfolded by
separating Hopf and pitchfork bifurcation or by lifting the degeneracy
of the pitchfork bifurcation, which we both demonstrate for our system
below. In this sense the Hopf-pitchfork bifurcation is a
codimension-three bifurcation.  Our results from analytical
considerations and numerical simulations are verified by measurements
of tunnel diode circuits.

This paper is organized as follows. In Sec.~\ref{sec:model}, the
circuit and the model will be introduced and compared with electronic
transport models for superlattices and with the FHN system. In
Sec.~\ref{sec:sin}, a single element will be investigated analytically
and numerically. Section~\ref{sec:two} deals with a circuit of two tunnel diodes in series, 
and Sec.~\ref{sec:arb} treats an
arbitrary number of tunnel diodes, where we find a generalization of
the Hopf-pitchfork bifurcations found in Sec.~\ref{sec:two} depending on the number of coupled
tunnel diode elements. In Sec.~\ref{sec:com}, measurements are compared to 
numerical simulations of the tunnel diode
circuits. Finally, conclusions are presented in Sec.~\ref{sec:con}.

\section{The circuit model}
\label{sec:model}

\begin{figure}[ht]
  \includegraphics[width=\textwidth]{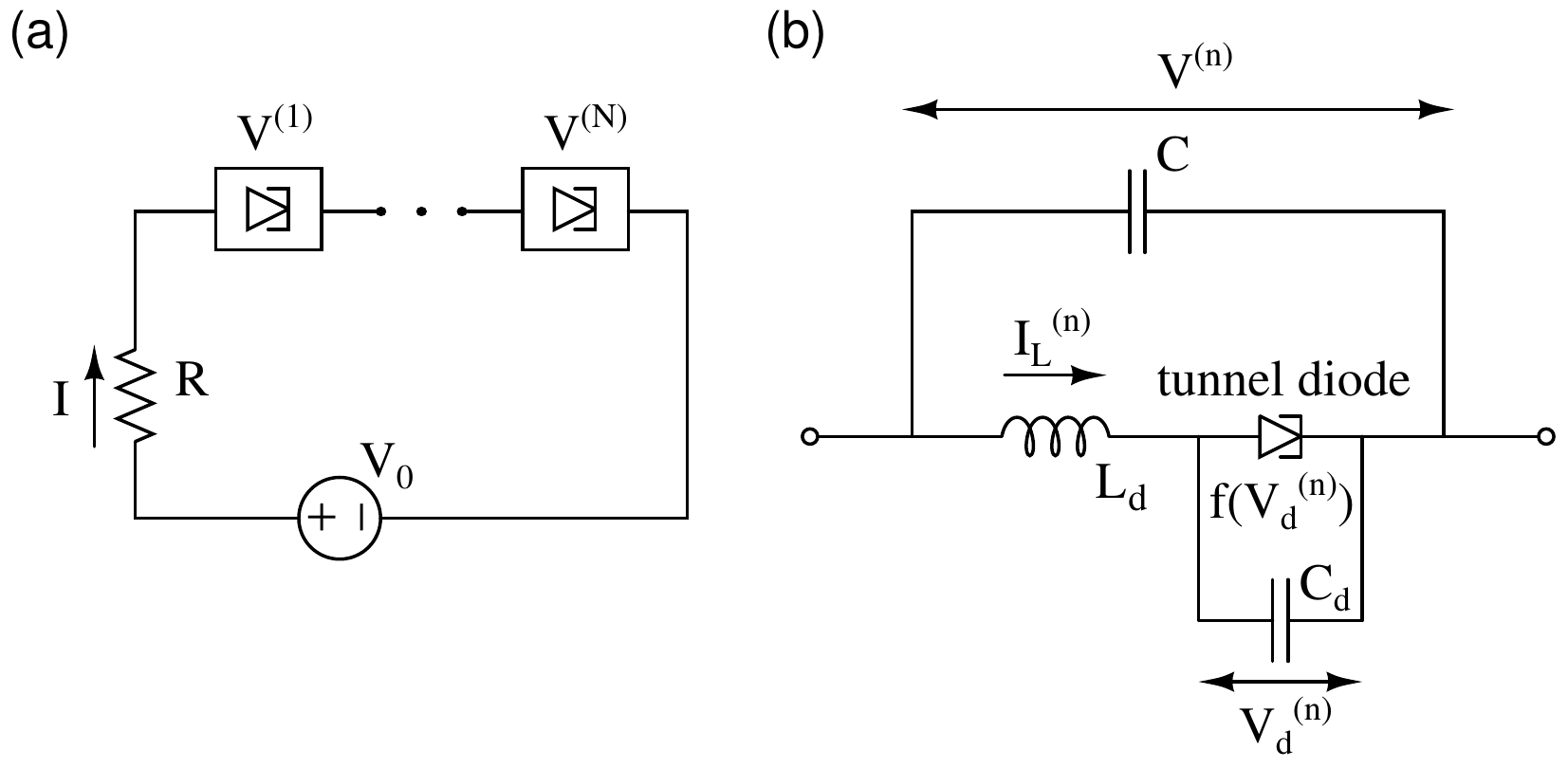}
  \caption{(a) Circuit built from $N$ tunnel diode elements
    according to panel (b) connected in series. $I$, $R$, and $V_0$
    denote the overall current, the load resistance, and the bias voltage,
    respectively. (b) Circuit diagram of a tunnel diode element including
    parallel capacitor $C$, intrinsic capacitance $C_d$, inductance
    $L_d$ with corresponding current $I_L^{(n)}$, and tunnel diode
    with the ideal current-voltage characteristic $f(V_d)$ according
    to Eq.~(\ref{eq:cubic}). $V_d^{(n)}$ and $V^{(n)}$ denote the
    voltages across the capacitor $C_d$ and across the whole element,
    respectively.}
  \label{img:Ntds}
\end{figure}

Figure~\ref{img:Ntds}(a) shows the circuit of $N$ tunnel diodes
connected in series. A single element, which contains the internal
device capacitance $C_d$ and inductance $L_d$ of the tunnel diode
\cite{SHA92,SCH00} as well as an external capacitor $C$, is shown in
Fig.~\ref{img:Ntds}(b). The superscript $n$ labels the $n$-th circuit element.

The current $f(V_d)$ represents the ideal tunnel diode $I$-$V$
characteristic, shown as solid black curve in Fig.~\ref{img:bascufi},
by neglecting any device capacitance and inductance. The $I$-$V$ curve
$f(V_d)$ of an ideal tunnel diode can be approximated by a cubic
polynomial function \cite{NAG62}
\begin{equation}
 \label{eq:cubic}
f(V_d)=I_0 - \frac{1}{\rho}\left( \left( V_{d} - V_{d_{0}} \right) -
  \frac{1}{3K^2}\left( V_{d} - V_{d_{0}} \right)^3\right),
\end{equation}
where $V_d$ and $f(V_d)$ are the voltage across and the current
through the tunnel diode, respectively. The parameters $V_{d_{0}}$ and $I_0$ are the
voltage and the current at the inflection point of the cubic function, respectively, $K$ is the
distance of $V_{d_{0}}$ to both extrema and $-1/\rho$ is the slope at $V_{d_{0}}$, as shown in
Fig.~\ref{img:bascufi}.

Coupling $N$ tunnel diode elements in series as shown in
Fig.~\ref{img:Ntds}(a) yields the following circuit equations using
Kirchhoff's laws: \numparts
\begin{eqnarray}
  V_0 &= R I + \sum^{N}_{n=1} V^{(n)}, \label{eq:kirch1} \\
  I &=  C^{(i)}\dot{V}^{(i)}+I_{L}^{(i)},  \label{eq:kirch2}\\
  V^{(i)} &= L_d^{(i)} \dot{I}_{L}^{(i)} + V_{d}^{(i)},  \label{eq:kirch3}\\
  I_{L}^{(i)} &= C_{d}^{(i)}\dot{V}_{d}^{(i)} + f^{(i)}(V_{d}^{(i)}),  \label{eq:kirch4}
\end{eqnarray}
\endnumparts
where $i = 1\, ,\, \ldots\, ,\,  N$. The variables $I$ and $I_L^{(i)}$ are the currents through the
resistor $R$ and through the inductor $L_d^{(i)}$, respectively, while
the variables $V^{(i)}$ and $V_d^{(i)}$ are the voltages across the
capacitors $C^{(i)}$ and $C_d^{(i)}$, respectively, and $V_0$ is the
bias voltage.

Assuming that all tunnel diode elements in Fig.~\ref{img:Ntds}(a) are
identical\footnote{An investigation of non-identical elements will be
  briefly covered in Sec.~\ref{sec:two}.} ($C^{(i)}=C$, $C_d^{(i)}=C_d$, $L_d^{(i)}=L_d$) and eliminating the current
$I$ in Eqs.~\eref{eq:kirch1} and \eref{eq:kirch2} yields the system of
dynamical equations for the variables $V^{(i)}$, $I_L^{(i)}$ and $V_d^{(i)}$ ($i = 1\, ,\,  \ldots\, ,\,  N$):\numparts
\begin{eqnarray}
  R C \dot{V}^{(i)} &=&  V_0 - \sum^{N}_{n=1} V^{(n)} -R I_{L}^{(i)},  \label{eq:modelunscaled1}\\
  L_d\dot{I}_L^{(i)} &=& V^{(i)} -V_d^{(i)}, \label{eq:modelunscaled2} \\
  C_d \dot{V}_d^{(i)} &=&  I_L^{(i)} - f(V_d^{(i)}). \label{eq:modelunscaled3}
\end{eqnarray}
\endnumparts

\subsection{Tunnel diodes as a superlattice model}
\label{sec:tunnel-diodes-as}

Let us discuss the similarities of superlattices and tunnel diodes by
deriving the regime of validity of Eqs.~\eref{eq:modelunscaled1}--\eref{eq:modelunscaled3}
as a model for the carrier dynamics in a superlattice following
Refs.~\cite{AMA04,AMA01,XU07}.

A superlattice consists of a sequence of quantum wells $i=1,\ldots,N$ and can be described by a system of differential equations
\begin{equation}
 \label{eq:superlattice}
  \epsilon \frac{\textnormal{d} F_i(t)}{\textnormal{d} t} + J_{i \rightarrow i+1}(F_i(t)) = J(t),
\end{equation}
for the electrical field strength $F_i$ at the $i$th potential
barrier. Here, $\epsilon$ is the permittivity and the term $\epsilon
\, \mathrm{d}F_i/\mathrm{d}t$ describes the displacement current
density.  The conduction current density $J_{i \rightarrow i+1}(F_i)$
between well $i$ and well $i+1$ is a tunnel current and can roughly be
approximated by a cubic polynomial as in Eq.~\eref{eq:cubic}. The sum
of the displacement current density and the conduction current density
gives the total current density $J$, which is independent of the well
index.

In the tunnel diode model the displacement current corresponds to the
current through the capacitor $C$ and the conduction current corresponds
to the current through the inductor $L_d$. The equation resulting from
Kirchhoff's laws is Eq.~\eref{eq:kirch2}.
In the approximation of vanishing device inductance ($L_d=0$) and
device capacitance ($C_d=0$) the current $I_L^{(i)}$ is given by the
current-voltage characteristic of the tunnel diode $ f(V^{(i)})$
(see Eqs.~\eref{eq:modelunscaled2} and \eref{eq:modelunscaled3}) and
Eqs.~\eref{eq:modelunscaled1}--\eref{eq:modelunscaled3} reduce to:
\begin{equation}
\label{eq:reducedmodelforsuperlattice2}
C\dot{V}^{(i)}+f(V^{(i)}) = I.
\end{equation}
The dynamical equation~\eref{eq:superlattice} describing a
superlattice is formally identical to the dynamical
system~\eref{eq:reducedmodelforsuperlattice2} describing a reduced
tunnel diode circuit. A more intuitive approach to explain the
similarity between both dynamical systems is to model each potential
barrier in a superlattice by a tunnel diode with a parallel capacitor,
which describes the charge accumulation in the quantum well, resulting
in building up the field in the barrier. The quantum wells are in this
case reduced to wires connecting the individual tunnel
diode--capacitor circuits to each other.

Equation~\eref{eq:reducedmodelforsuperlattice2} is only an
approximation of the carrier dynamics in superlattices since it uses
only a very rough approximation of the conduction current density
$J_{i \rightarrow i+1}(F_i)$. The conduction current density
introduces an additional local coupling besides the global coupling
through the circuit in Eqs.~\eref{eq:superlattice} and
\eref{eq:reducedmodelforsuperlattice2}, since $J_{i \rightarrow
  i+1}(F_i)$ depends also upon the carrier densities $n_i$, $n_{i+1}$
in the neighboring quantum wells, which are in turn coupled to the
fields $F_i$ by Gauss' law \cite{SCH00}. This additional
nearest-neighbor coupling cannot be observed in the tunnel diode
model. Therefore, spatio-temporal patterns like high and low field
domains cannot be modeled by the tunnel diode circuit.

\subsection{Rescaling the model}
\label{sec:rescaling-model}

In the following, a scaling is introduced to simplify the circuit
model~\eref{eq:modelunscaled1}--\eref{eq:modelunscaled3} for all
numerical and analytical investigations in this paper. The scaling
combines circuit parameters and eliminates the physical dimensions
from the dynamical variables so that a set of only four independent
dimensionless parameters remains for the analysis. Using this scaling,
the analogy of the tunnel diode with the FitzHugh-Nagumo model, which
is widely used to model neuronal dynamics, becomes more apparent.

Considering identical tunnel diode elements, we apply the following
scaling. The dimensionless variables and parameters are given by
\begin{eqnarray}
  x_i(t) &=&  - \frac{1}{K} (V_{d}^{(i)}(\tau) - V_{d_0}), \label{eq:scaling} \\
  y_i(t)&=&  \frac{\rho}{K}(I_{L}^{(i)}(\tau) - I_0),\nonumber\\
  z_i(t) &=&  \frac{1}{K} ( V^{(i)}(\tau) - V_{d_0}),\nonumber\\
  t &=&  \frac{\rho}{L}\tau,\nonumber \\
  \epsilon &=& \frac{\rho^2 C_d}{L},\nonumber\\
  \gamma &=& \frac{RC \rho}{L},\nonumber\\
  d&=&\frac{R}{\rho},\nonumber\\
  a^{(N)} &=&  \frac{1}{K}\left(V_0 - N V_{d_0}-RI_0\right).\nonumber
\end{eqnarray}
With these new variables, the
system~\eref{eq:modelunscaled1}--\eref{eq:modelunscaled3} simplifies to \numparts
\begin{eqnarray}
  \epsilon \dot{x}_i &=  x_i - \frac{1}{3} x_i^3 - y_i, \label{eq:modela}\\
  \dot{y}_i &=  x_i +  z_i, \label{eq:modelb} \\
  \gamma \dot{z}_i &=  a^{(N)} - \sum^{N}_{n=1} z_n -dy_i. \label{eq:modelc}
\end{eqnarray}
\endnumparts
Only the four dimensionless parameters $a^{(N)}$, $d$, $\epsilon$, and
$ \gamma$ remain in the scaled equations, where $\epsilon$ and
$\gamma$ are timescale parameters, given by the reactive circuit
components.  The parameter $d$ is given by the ratio of slopes of the
load line and the tunnel characteristics at the inflection point (see
Fig.~\ref{img:bascufi}) and will be limited to $0 < d < 1$, such that
there is a single intersection of the load line and the tunnel diode
$I$-$V$ curve, resulting in a single fixed point. The parameter
$a^{(N)}$ depends on the number $N$ of elements and contains the bias
voltage $V_0$.  The calculations in this work are performed with
dimensionless variables, but for better comparison with experimental
data all results are presented in terms of the circuit variables
$V_d^{(i)}$, $I_L^{(i)}$ and $V^{(i)}$, corresponding to $x_i$, $y_i$
and $z_i$, respectively.

In Tab.~\ref{tab:comsin1} the circuit parameters of the measurements
and simulations are listed. The simulation parameters are obtained by
using the best possible fit of the measured data, by considering the
fabrication variance of the devices. This leads to the small
difference in parameters for $C$ and $R$. In Tab.~\ref{tab:2} the
physical units of the dimensionless variables derived from the
parameters are given.

\begin{table}[htb]
 \centering
\begin{tabular}{cccccccc}
 \toprule
$V_{d_{0}}$   & $K$  & $I_0$  & $\rho$ & $L$   & $C_d$  & $C$  & $R$\\
\midrule
0.267 V & 0.1306 V & 1.22 mA & 86 $\Omega$ &10 nH & 26 pF & 56(55) pF & 47(49) $\Omega$\\
\bottomrule
\end{tabular}
\caption{Parameters of the tunnel diodes as measured in the
  experiments and used in the simulations (in brackets where different).}
\label{tab:comsin1}
\end{table}

\begin{table}[htb]
 \centering
\begin{tabular}{rl|rll}
 \toprule
$\epsilon$&$= 19.23$   & $\tau $&$= t L/\rho $&$= (0.116 t) \, \mathrm{ns}$\\
$\gamma $&$= 23.18$ & $V_d $&$= -Kx + V_{d_0} $&$= (-0.1306 x + 0.267)
\, \mathrm{V}$ \\
$d $&$= 0.57$ & $I_L$&$=\frac{K}{\rho}y + I_0 $&$= (1.52 y + 1.22) \, \mathrm{mA}$\\ 
$a^{(1)} $&$= 7.657 V_0/V -2.502$ & $V $&$= Kz + V_{d_0} $&$= (0.1306 z
+ 0.267) \, \mathrm{V}$\\
$a^{(N)} $&$= 7.657 V_0/V - 2.044 N - 0.4578$ & &\\
\bottomrule
\end{tabular}
\caption{Scaled parameters and variables derived from
  Tab.~\ref{tab:comsin1} using the scaling~(\ref{eq:scaling}).}
\label{tab:2}
\end{table}

A tunnel diode modeled by Eqs.~\eref{eq:modela}--\eref{eq:modelc} reduces
to the FHN model \cite{FIT61,NAG62,IZH07} for $R=0$, which leads to $\gamma=0$ and
$d=0$. In the case of a single element ($N=1$), the parameter
$a^{(1)}$ is then equal to $z$. The resulting system reads \numparts
\begin{eqnarray} 
  \epsilon \dot{x} &=  x - \frac{1}{3} x^3 - y, \label{eq:modelfitzhugh1}\\
  \dot{y} &=  x +  a^{(1)}. \label{eq:modelfitzhugh2}
\end{eqnarray}
\endnumparts 
These equations are of the same form as the simplified FHN system,
widely used in neurodynamics, e.g.,
\cite{JAN03,SCH08,DAH08c,SCH09c,HOE09,BRA09}. Note that the values
of the parameters $\epsilon$ and $\gamma$ we have
used to describe the experiment differ from the ones commonly used in
the FHN system. We stress, however, that the location of the fixed
points is independent of these parameters. Furthermore, for other
circuit parameters small $\epsilon$ and $\gamma$ can be realized,
resembling the FHN model.

\section{A single tunnel diode}
\label{sec:sin}

First, we perform a bifurcation analysis of a single tunnel diode. The
dynamics of a single diode resembles the synchronized dynamics of $N$
tunnel diodes in series, however with a different parameter $a^{(N)}$,
see Eqs.~(\ref{eq:scaling}).

The fixed point ($x^*$, $y^*$, $z^*$) of
Eqs.~\eref{eq:modela}--\eref{eq:modelc} for $N=1$ is given by a third
order polynomial for the $z$ component:
\begin{equation}
  \frac{1}{3}(z^*)^3 +\left(\frac{1}{d}-1\right)z^*-\frac{a^{(1)}}{d}=0.  \label{eq:fixpsingle}
\end{equation}
and $x^*=-z^*$, $y^*=(a-z^*)/d$. The fixed point depends on the parameter $a^{(1)}$, which includes the applied voltage
$V_0$, and on the ratio $d=R/\rho$ (see Eqs.(\ref{eq:scaling})). To
calculate the stability of the fixed point, we consider the
eigenvalues of the Jacobian
\begin{equation}
\label{eq:jacsingle}
 \mathbf{J}^{(1)}=  \left( \begin{array}{ccc} \frac{1}{\epsilon}(1 - (x^*)^2) & -\frac{1}{\epsilon} & 0 \\ 1 & 0 & 1 \\ 0 & -\frac{d}{\gamma} &-\frac{1}{\gamma} \end{array} \right)
\end{equation}
evaluated at the fixed point. The stability of the fixed point depends on $x^*(d,a^{(1)})$, as well as on the timescale parameters
$\epsilon$ and $\gamma$ and the parameter $d$.

\begin{figure}[ht]
 \includegraphics[width=\textwidth]{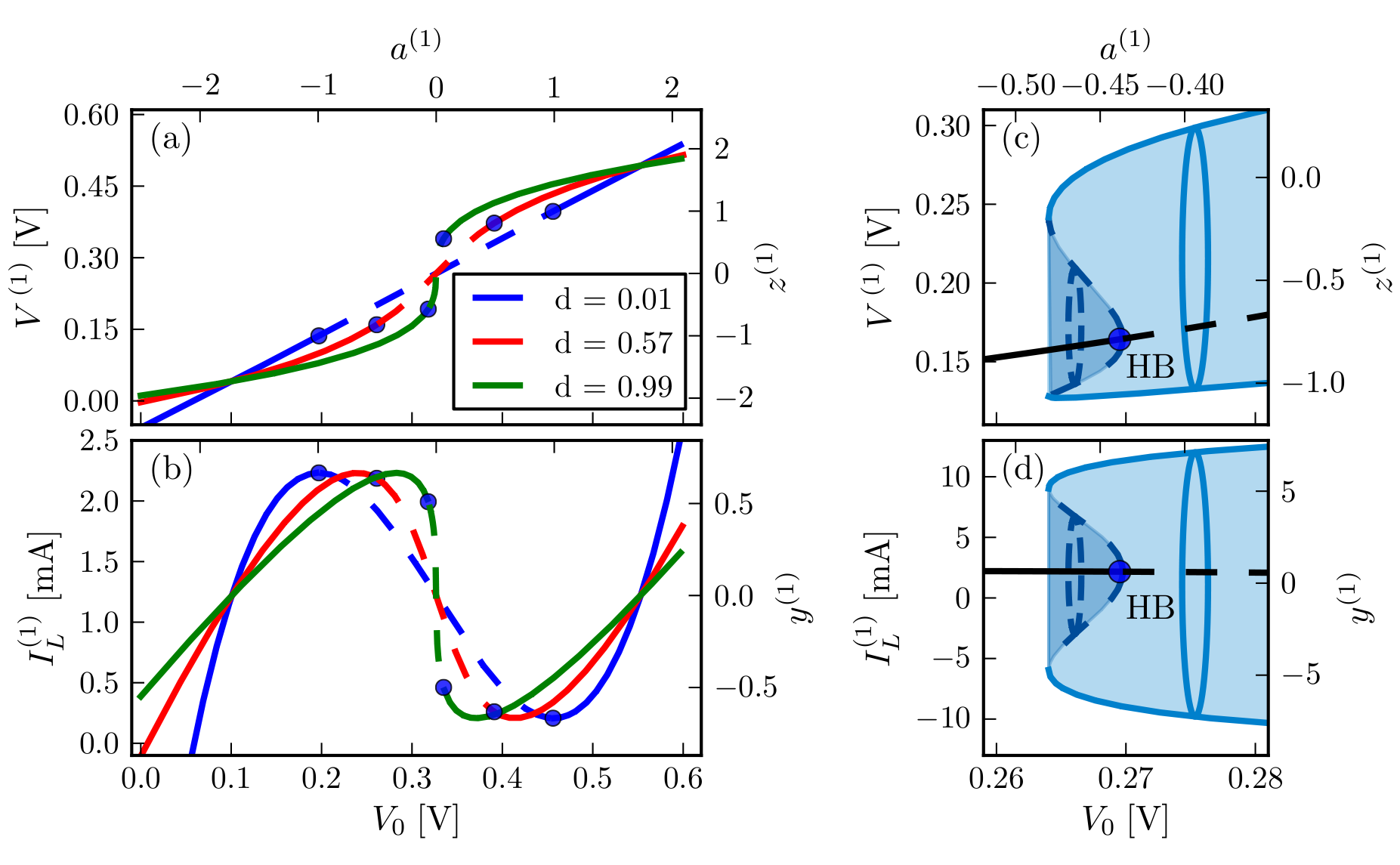}
 \caption{Voltage $V^{(1)}$ (a) and current $I_L^{(1)}$ (b) at the
   fixed points according to Eq.~\eref{eq:fixpsingle} for a single
   element vs.~the bias voltage $V_0$ for different values of $d$. The
   blue, red, and green lines correspond to $d=0.01$, $d=0.57$, and
   $d=0.99$, respectively. Blue dots mark subcritical Hopf
   bifurcations (HB). Panels (c) and (d) show fixed points and
   periodic orbits for $d = 0.57$ in voltage ($V^{(1)}$) and current
   ($I^{(1)}_L$), respectively. The black solid and dashed lines
   correspond to stable and unstable fixed points, respectively. The
   solid light blue and dashed dark blue lines correspond to minimum
   and maximum values of stable and unstable periodic orbits,
   respectively. Color shading shows location and overlap of these
   periodic orbits. A stable and an unstable orbit are depicted
   symbolically for better visualization. Parameters: $\epsilon =
   19.23$, $\gamma=23.18$, $a^{(1)}=7.657 V_0-2.502$. Movie 1 shows
   the subcritical Hopf bifurcation additionally in three dimensional
   phase space.}
 \label{img:fixsingle}
\end{figure}

The $V^{(1)}$- and $I_L^{(1)}$-values (and the corresponding
$z^{(1)}$- and $y^{(1)}$-values) of the fixed point are shown in
Figs.~\ref{img:fixsingle}(a) and (b) as a function of the bias $V_0$
for different values of the parameter $d$, where the blue, red, and
green lines denote $d=0.01$, $d=0.57$, and $d=0.99$,
respectively. Solid and dashed lines correspond to stable and unstable
fixed points, respectively. For $d$ close to zero ($d=0.01$, blue
line), it follows from Eq.~\eref{eq:fixpsingle} that $z^*\approx
a^{(1)}$, and the voltage curve $V^{(1)}(V_0)$ is approximately a
straight line. This case occurs for example if the load resistor $R$
is very small and corresponds to the limit of the FHN system in
Eqs.~\eref{eq:modelfitzhugh1}--\eref{eq:modelfitzhugh2}.

Figures~\ref{img:fixsingle}(a) and (b) also show that the location of
the fixed point exhibits a point symmetry around $(V_0,\, V^{(1)}) =
(0.327\,\mathrm{V}, 0.267\,\mathrm{V})$, which is equivalent to
$a^{(1)} = 0$ and $z_1=0$. This is caused by the point symmetry of the
cubic polynomial in Eq.~\eref{eq:fixpsingle}.

For each curve, i.e., each value of $d$, the fixed point is unstable
in a specific interval of $V_0$. Using the eigenvalues of the Jacobian
matrix and numerical bifurcation analysis we will show that the fixed
point changes its stability at the boundaries of this interval in
subcritical Hopf bifurcations.

Figures~\ref{img:fixsingle}(c) and (d) show a bifurcation diagram at
the first subcritical Hopf bifurcation (low $V_0$) for $d=0.57$. The
plots were obtained using the continuation tool \texttt{Auto}
\cite{DOE09}. The light blue (stable) and dashed dark blue (unstable)
lines show the maxima and minima of $V^{(1)}$ (c) and $I_L^{(1)}$ (d)
of the periodic orbit. For better visualization an ellipse is plotted
for each branch of stable or unstable periodic orbits. The position where
the subcritical Hopf bifurcation (HB) occurs is denoted by the blue
dot at $V_0=0.2698\,\mathrm{V}$. At $V_0=0.2640\,\mathrm{V}$ the
unstable and stable limit cycles collide in a saddle-node bifurcation
of limit cycles. An illustration of the subcritical Hopf bifurcation
in three dimensional phase space can be found in movie 1 in the
supplementary material.

For very small values of the resistance $R$, i.e., in the limit of the
simplified FHN system, the Hopf bifurcations occur at the extrema of
the current-voltage curve (blue curve in
Fig.~\ref{img:fixsingle}(b)). This shows that the tunnel diode system
gives a smooth approach to the simplified FHN system.

In the following sections we will show that the synchronized state of
$N$ tunnel diodes shows similarities to a single tunnel diode but with
a stretching factor $N$ as introduced in parameter $a^{(N)}$ in
Eqs.~\ref{eq:scaling}.

\section{Two tunnel diodes in series}
\label{sec:two}

In this section we perform a bifurcation analysis of two tunnel diodes
connected in series. We see that a codimension-three Hopf-pitchfork
bifurcation leads to a symmetry-breaking transition. Introducing
heterogeneity to the system of tunnel diodes lifts the degeneracy and
unfolds the Hopf-pitchfork bifurcation.

\subsection{Identical tunnel diodes}
\label{sec:twoident}
Similar to the analysis of a single element in Sec.~\ref{sec:sin}, we
can calculate the fixed points in the case of $N=2$ in
Eqs.~\eref{eq:modela}--\eref{eq:modelc}. We find two coupled third
order polynomial equations for $z_1^*$ and $z_2^*$ of the fixed point
($x_1^*$, $y_1^*$, $z_1^*$, $x_2^*$, $y_2^*$, $z_2^*$): \numparts
\begin{eqnarray} 
\frac{1}{3} (z^*_{1})^3-z^*_{1}-\frac{1}{d}(a^{(2)}-z^*_{1}- z^*_{2}) &= 0 \label{eq:fixtwo1}\\
\frac{1}{3} (z^*_{2})^3-z^*_{2}-\frac{1}{d}(a^{(2)}-z^*_{1}- z^*_{2})  &= 0 . \label{eq:fixtwo2}
\end{eqnarray}
\endnumparts 
Again both equations depend on $a^{(2)}$ ($V_0$) and on the ratio
$d=R/\rho$. The Jacobian at the fixed points is given by:
\begin{equation}\fl
\label{eq:jacstwo}
  \mathbf{J}^{(2)}=\left( \begin{array}{cccccc}  1/\epsilon(1-(x^*_{1})^2) & -1/\epsilon &0 &0&0&0\\  1 & 0 & 1 &0&0&0\\  0& -d/\gamma&-1/\gamma&0&0&-1/\gamma \\0&0&0&1/\epsilon(1-(x^*_{2})^2) & -1/\epsilon&0\\0&0&0& 1 & 0 & 1\\0&0&-1/\gamma&0&-d/\gamma&-1/\gamma \end{array} \right),
\end{equation}
with $x^*_1 = -z^*_1$ and $x^*_2 = -z^*_2$. The stability of the fixed
points depends on the $x$ component of both fixed points, as well as
the timescale parameters $\epsilon$ and $\gamma$ and parameter $d$. 

\begin{figure}[ht]
  \includegraphics[width=\textwidth]{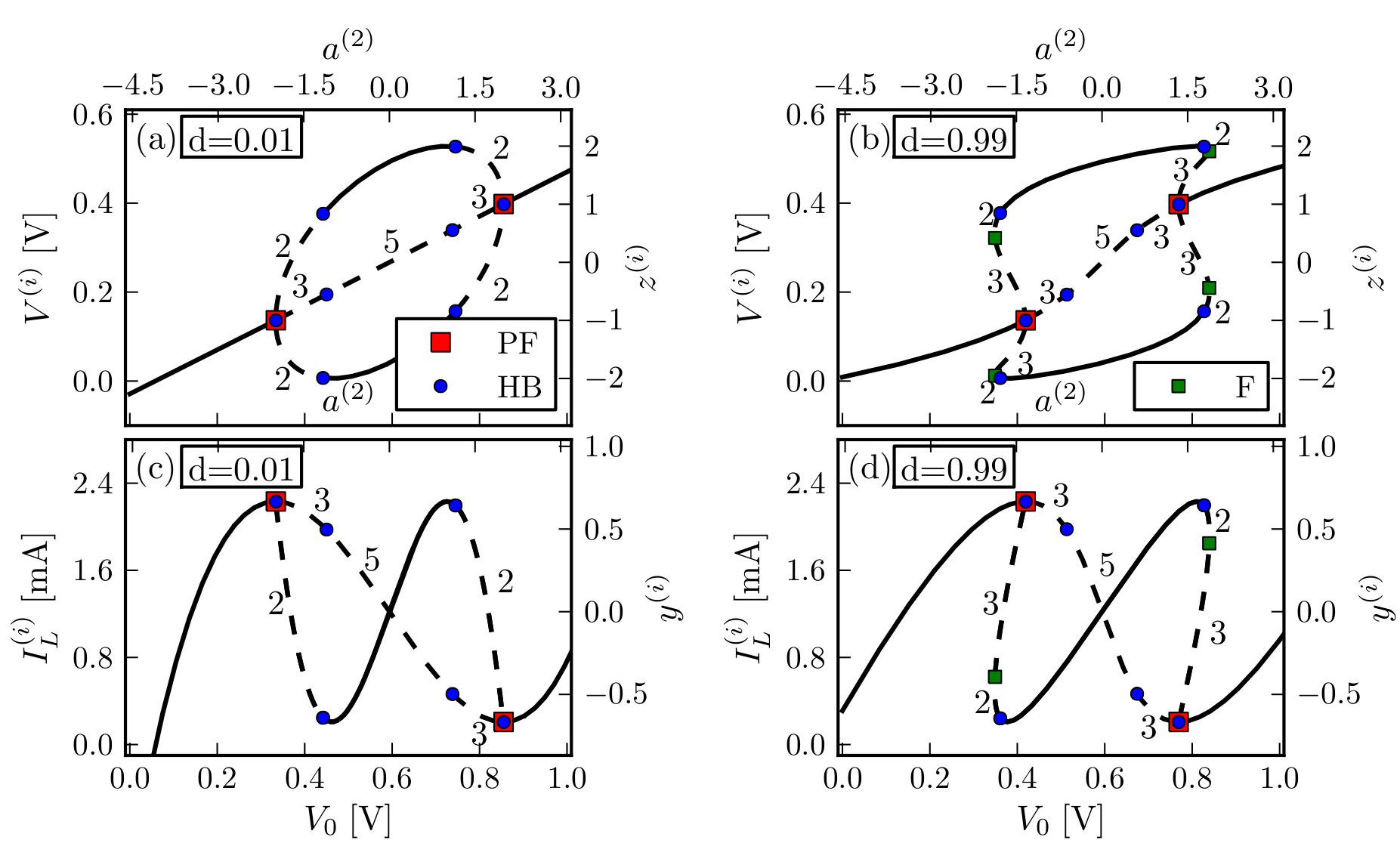}
  \caption{Voltage $V^{(i)}$ (panels (a) and (b)) and current
    $I_L^{(i)}$ (panels (c) and (d)) of the calculated fixed points of
    two tunnel diode elements according to
    Eqs.~(\eref{eq:modela}-\eref{eq:modelc}) vs.~$V_0$. Solid lines
    denote stable fixed points, while dashed lines are unstable fixed
    points. The labels on the branches denote the unstable
    dimension. Panels (a) and (c) correspond to $d=0.01$ and panels
    (b) and (d) correspond to $d=0.99$. Red squares and blue dots mark
    the location of pitchfork and Hopf bifurcations,
    respectively. Green dots are fold bifurcations (F).  Note that the
    pitchfork bifurcations coincide with Hopf bifurcations and the
    number of unstable dimensions changes by three at these degenerate
    points.  Parameters: $\epsilon = 19.23$, $\gamma=23.18$,
    $a^{(2)}=7.657 V_0-4.547$.}
  \label{img:fix}
\end{figure}

Figure~\ref{img:fix} shows the $V^{(i)}$- and the $I_L^{(i)}$-values
(as well as the corresponding $z^{(i)}$- and $y^{(i)}$-values) of
the fixed points for two tunnel diode elements as a function of
$V_0$. Panels (a) and (c) correspond to $d=0.01$, and panels (b) and
(d) correspond to $d=0.99$. Solid and dashed lines again correspond to
stable and unstable branches, respectively. The number next to each
unstable branch indicates the number of positive eigenvalues and thus
the number of unstable dimensions.

The red squares and blue dots in Fig.~\ref{img:fix} mark pitchfork
(PF) and Hopf bifurcations (HB), respectively. The green squares in
(b) and (d) mark fold bifurcations (F), where two fixed points
(saddles) with unstable dimension $2$ and $3$ collide.

The point symmetry discussed above for a single element (refer to
Sec.~\ref{sec:sin}) remains valid for the system of two tunnel
diodes. But the symmetry point is shifted to $(V_0,\, V) =
(0.594\mathrm{V}, 0.267\mathrm{V})$ due to the rescaled value of
$a^{(N)}$.

Another symmetry in the two-diode system is a $Z_2$ symmetry
corresponding to the interchange of the dynamical variables of tunnel
diode one and two, i.e., the Eqs.~\eref{eq:fixtwo1} and
\eref{eq:fixtwo2} and \eref{eq:modela}--\eref{eq:modelc} are invariant
under the transformation
\begin{equation} 
  (x_1,\, y_1,\, z_1,\, x_2,\, y_2,\, z_2) \mapsto (x_2,\, y_2,\, z_2,\, x_1,\, y_1,\, z_1) \label{eq:symmetry-trans}
\end{equation}
and the phase space is symmetric with respect to reflections at the
synchronization manifold.\footnote{As introduced in \cite{PEC97} the
  synchronization manifold is a hyperplane in phase space of a system,
  where the synchronized dynamics take place.} The synchronization
manifold $(x_1,y_1,z_1)=(x_2,y_2,z_2)$ shows identical behavior to the
dynamics of a single tunnel diode only with a rescaled parameter
$a^{(N)}$ as introduced in Eqs.~\eref{eq:scaling}. All fixed points
represented by the middle branch in Fig.~\ref{img:fix} are located
within the synchronization manifold, since both tunnel diodes voltages
$V^{(1)}$ and $V^{(2)}$ are the same.

The pitchfork bifurcations at $V_0 = 0.334\mathrm{V}$ and
$V_0=0.855\mathrm{V}$ for $d=0.01$ and at $V_0 = 0.419\mathrm{V}$ and
$V_0 = 0.619\mathrm{V}$ for $d=0.99$ mark a symmetry-breaking
transition: the synchronized state becomes unstable. Note that a Hopf
bifurcation coincides with the pitchfork bifurcation, thus the two
emerging fixed point branches are unstable and the system will
actually follow one of two stable periodic orbits, as we will show
below.

The pitchfork bifurcation can be super- or subcritical depending on
the parameter $d$ as shown in Fig.~\ref{img:fix}. The change from a
subcritical pitchfork bifurcation to a supercritical pitchfork
bifurcation occurs at $d =0.34$ for the parameter set in
Fig.~\ref{img:fix}.

For all values of $d$ between zero and one, a Hopf bifurcation occurs
simultaneously with the pitchfork bifurcations, i.e., a complex
conjugate pair of eigenvalues as well as a real eigenvalue cross the
imaginary axis in the complex plane simultaneously. To find the
location of this Hopf-pitchfork bifurcation further, we proceed as
follows: The characteristic polynomial $\chi(\lambda)$ is of $6$th
order and thus difficult to solve directly.  However, below the
Hopf-pitchfork bifurcation there is only one symmetric fixed point,
i.e., $x \equiv x_1=x_2$. Furthermore, at the pitchfork bifurcation we
have $\chi(0) \stackrel{!}{=} 0$ which gives (after calculating the
characteristic polynomial)
\begin{equation*} 
  0 = d^2(1-x^2)^2 - 2d(1-x^2).
\end{equation*}
This equation has four solutions for $x$
\begin{equation*} 
  x = \pm 1,\qquad x=\pm\sqrt{\frac{d-2}{d}}.
\end{equation*}
Since we consider $0<d<1$, the latter two solutions are imaginary and
thus spurious and we have found the coordinate values $x=\pm 1$ at the
Hopf-pitchfork bifurcations.

The fixed point equations \eref{eq:fixtwo1} and \eref{eq:fixtwo2}
reduce for $x=-z_1=-z_2$ to
\begin{equation*} 
  0 = \frac{1}{3}x^3 + (\frac{2}{d}-1)x +\frac{a^{(2)}}{d}.
\end{equation*}
Inserting now $x=\pm 1$ we find the location of the Hopf-pitchfork bifurcation
\begin{equation*} 
 a^{(2)} = \pm \frac{d}{3} \pm (2-d).
\end{equation*}
In \ref{sec:appendixa}, we analyze the characteristic polynomial in
the general case of $N$ tunnel diodes.

\begin{figure}[ht]
 \includegraphics[width=\textwidth]{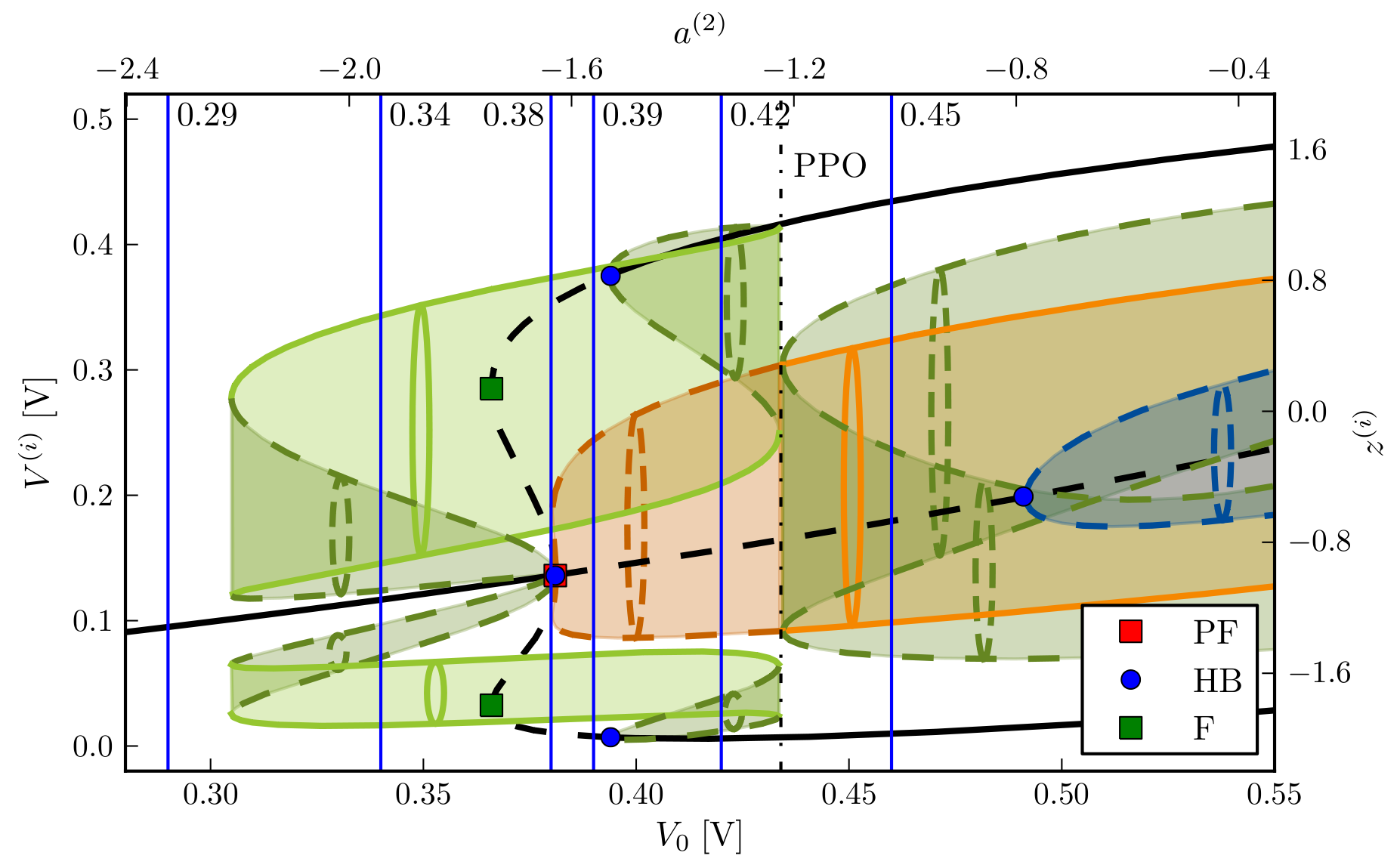}
 \caption{Bifurcation diagram. Voltage ($V^{(i)}$) coordinate of the
   fixed points and periodic orbits of the system vs.~the
   bias voltage $V_0$.
   Black solid and black dashed lines denote stable and unstable fixed
   points, respectively.  Green, orange, and blue lines correspond to
   the maxima and minima of periodic skew, anti-phase, and in phase
   orbits, respectively, where solid and dashed lines denote stable
   and unstable periodic orbits, respectively.  Color shading shows
   location and overlap of these periodic orbits. Limit cycles are
   plotted symbolically as ellipses for better visualization. PPO is a
   Pitchfork bifurcation of periodic orbits.  $d=0.57$, other
   parameters as in Fig.~\ref{img:fix}. Movie 2 shows the
   Hopf-pitchfork bifurcation in the six-dimensional phase
   space. Movies 3 and 4 show examples for skew and anti-phase
   oscillations, respectively.}
  \label{img:autbo}
\end{figure}

Figure~\ref{img:autbo} shows a bifurcation diagram at the first
subcritical pitchfork bifurcation (low $V_0$) for $d=0.57$ obtained
with the continuation tool \texttt{Auto}. Again, solid black lines
represent stable fixed points, while dashed black lines represent
unstable fixed points. The bifurcation diagram shows possible
solutions for one of the tunnel diodes only. If one of the tunnel
diodes is located on the middle (symmetric) fixed point, the other
tunnel diode is also in this fixed point. If one of the tunnel diodes
is located on the upper branch of the fixed points, the other tunnel
diode is on the lower branch and vice versa.

The green, orange, and blue lines show the
maxima and minima of periodic orbits, where solid lines and light
color shading denote stable periodic orbits, and dashed lines and dark
color shading denote unstable periodic orbits. For better
visualization, an ellipse is plotted for each branch of stable or
unstable periodic orbits. We will discuss the differences between the
green, orange, and blue orbits below.

In Fig.~\ref{img:autbo} the Hopf-pitchfork bifurcation occurs at $V_0
= 0.381 \mathrm{V} $. Additionally, at each fixed point branch of the
pitchfork bifurcation another Hopf bifurcation occurs for larger
values of $V_0$: at the two outer branches at $V_0 = 0.394 \mathrm{V}$
and at the inner branch at $V_0 = 0.492 \mathrm{V}$. 

The phase space symmetry Eq.~\eref{eq:symmetry-trans} implies that
there can be three different types of periodic orbits, which we call
synchronized orbits, anti-phase orbits, and skew orbits, respectively. The
synchronized orbits (shown in blue) lie in the synchronization
manifold and thus obey
\begin{equation*} 
  (x_1(t),\, y_1(t),\, z_1(t)) = (x_2(t),\, y_2(t),\, z_2(t)).
\end{equation*}
The anti-phase orbits (orange) wind around the $3$-dimensional synchronization
manifold in the $6$-dimensional phase space and obey the strict
anti-phase relation
\begin{equation*} 
  (x_1(t),\, y_1(t),\, z_1(t)) = (x_2(t-T/2),\, y_2(t-T/2),\, z_2(t-T/2)),
\end{equation*}
where $T$ is the period of the orbit. Finally, the skew orbits (shown in green) are neither anti-phase nor synchronized and come in pairs, due to the
phase-space symmetry Eq.~\eref{eq:symmetry-trans}. They are similar to
the outer pitchfork branches for fixed points, i.e., one of the diode
voltages oscillates around a high voltage and the other around a low
voltage state.

At $V_0 = 0.434 \mathrm{V}$ the anti-phase orbit surrounding the inner
unstable fixed point becomes stable in a pitchfork bifurcation of
periodic orbits (PPO) and generates a pair of unstable skew orbits.
As is apparent from this bifurcation the skew orbits (at least close
to the pitchfork bifurcation) show a generalized anti-phase dynamics,
i.e., they do not have a strict anti-phase relation such as the
anti-phase orbits, but when there is a maximum in the time-series of
the high voltage diode, there is a minimum in the time-series of the
low voltage diode and vice versa.  In our numerical simulations we see
this behavior for all skew orbits and all parameters. We
illustrate this behavior in movie 3 which
can be found in the supplementary material.

The periodic orbits surrounding the middle fixed point branch in
Fig.~\ref{img:autbo} are strict anti-phase orbits and both tunnel
diode voltages oscillate in complete anti-phase. The supplementary
movie 4 illustrates this behavior.

\begin{figure}[ht]
  \includegraphics[width=\textwidth]{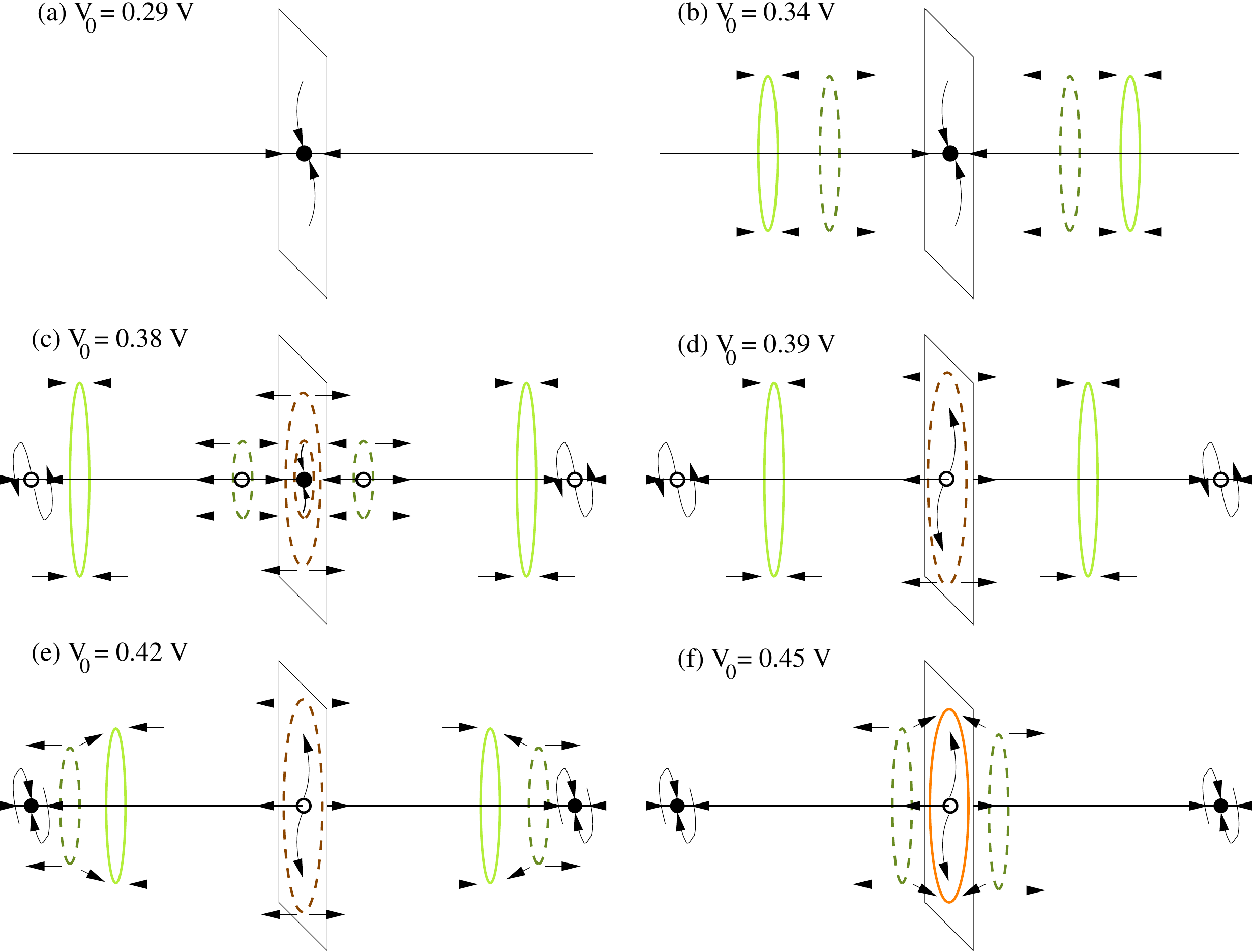}
  \caption{Schematic three-dimensional projection of the
    six-dimensional phase space showing the bifurcation mechanism as
    in Fig.~\ref{img:autbo}. Arrows indicate the direction of stable
    or unstable manifolds and thus the stability of the limit cycles
    and the fixed points. Panels (a), (b), (c), (d), (e), and (f)
    correspond to the bias voltages $V_0=0.29$~V, $0.34$~V, $0.38$~V,
    $0.39$~V, $0.42$~V, and $0.45$~V marked by blue lines in
    Fig.~\ref{img:autbo}. (a) a single stable fixed point, (b) two
    pairs of stable and unstable limit cycles are born in saddle-node
    bifurcations, (c) two unstable limit cycles are born in a fold
    bifurcation and and two unstable fixed points are born in a fold
    bifurcation on each side of the middle stable fixed point, (d)
    three unstable inner limit cycles and both unstable fixed points
    collapse with the middle fixed point in a Hopf-pitchfork
    bifurcation, (e) Hopf bifurcations of outer fixed points, (f)
    saddle-node bifurcation of outer limit cycles and pitchfork
    bifurcation of inner limit cycle.}
  \label{img:schetwo}
\end{figure}

Figure~\ref{img:schetwo} is a schematic three-dimensional projection
of the six-dimensional phase space showing the codimension-three
Hopf-pitchfork bifurcation, with values of $V_0$ as indicated in
Fig.~\ref{img:autbo} as blue lines. The arrows illustrate the
direction of the stable and unstable manifolds and thus the stability
of the limit cycles and fixed points. An illustration of the complete
bifurcation scenario shown in Fig.~\ref{img:autbo} can be found in the
supplementary material in movie 2.  The investigation of the
Hopf-pitchfork bifurcation by a schematic projection follows the same
spirit as the figures in \cite{KEH09}, where the stabilization of
complex spatio-temporal dynamics near a subcritical Hopf bifurcation
is investigated by time-delayed feedback.

In Fig.~\ref{img:schetwo}(a) a voltage of $V_0=0.29 \mathrm{V}$ is
applied. Only one stable fixed point exists. At $V_0=0.34 \mathrm{V}$,
as in Fig.~\ref{img:schetwo}(b), a stable and an unstable skew orbit
are born in a saddle-node (fold) bifurcation on each side of the
middle stable fixed point. The stable skew orbits move away from the
stable fixed point while the unstable skew orbits move towards the
fixed point with increasing $V_0$.

In Fig.~\ref{img:schetwo}(c) at $V_0=0.38 \mathrm{V}$, which is just
below the value of the Hopf-pitchfork bifurcation, two fixed points
are born in a fold bifurcation on each side of the middle
stable fixed point: one unstable fixed point and a saddle point, which
is stable in the direction of the pitchfork bifurcation but unstable
in the transversal direction. Also, a fold bifurcation of
anti-phase orbits takes place around the stable middle fixed
point. Both limit cycles are unstable in the direction of the
pitchfork bifurcation.

At the Hopf-pitchfork bifurcation two unstable skew orbits and one of
the unstable anti-phase orbits as well as both unstable fixed points
close to the center fixed point coalesce with the latter. The stable
fixed point in the center becomes
unstable. Figure~\ref{img:schetwo}(d) illustrates this
schematically. The outer two stable skew orbits and two unstable fixed
points as well as the unstable anti-phase orbit remain unchanged.

In Fig.~\ref{img:schetwo}(e), at $V_0=0.42 \mathrm{V}$, the two outer,
unstable fixed points have become stable in a subcritical Hopf
bifurcation by giving birth to two unstable skew orbits. These orbits
are moving towards the stable skew orbits which themselves are moving
away from the unstable fixed point in the center.

Finally, in Fig.~\ref{img:schetwo}(f) at $V_0=0.45 \mathrm{V}$ the two
outer unstable skew orbits collide with the two stable skew orbits in
a saddle-node bifurcation and the limit cycle around the centered
fixed point becomes stable in a PPO bifurcation and gives rise to two
unstable skew orbits.

\subsection{Introducing heterogeneity}
So far both tunnel diodes were assumed to be identical as assumed in
Eqs.~\eref{eq:modela}-\eref{eq:modelc}. This assumption is not
justified in experimental situations (see Sec.~\ref{sec:com}), since
it is impossible to find two perfectly identical electronic
devices. In what follows we investigate nonidentical devices by
introducing heterogeneity to the parameters of the subsystem. The
heterogeneity may be introduced as discrepancy in the parameter
$a^{(2)}$ or as discrepancy in one of the time scale parameters
$\epsilon$ or $\gamma$ in Eqs.~\eref{eq:modela}-\eref{eq:modelc}:

\begin{figure}[ht]
  \includegraphics[width=\textwidth]{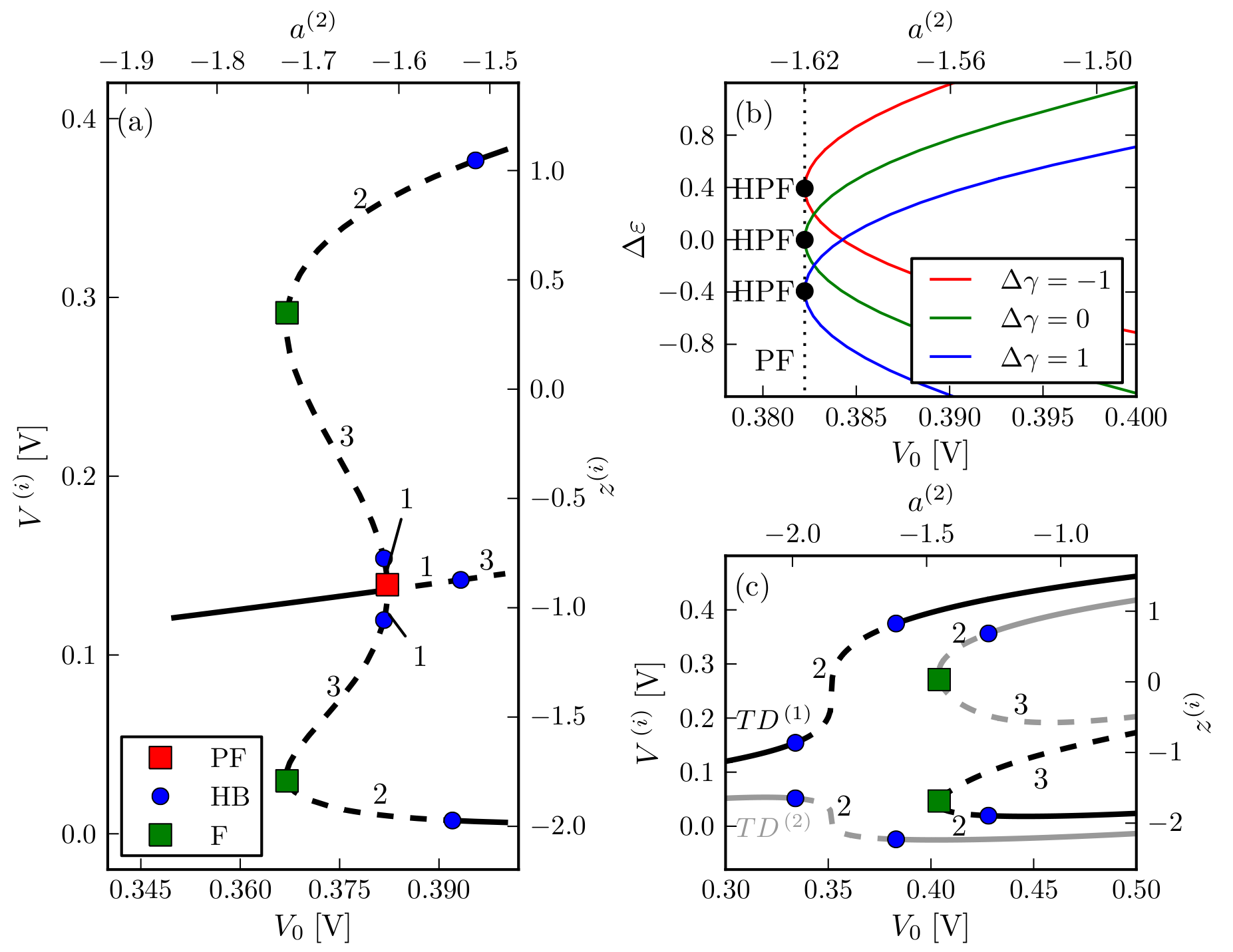}
  \caption{Fixed points and their stability for two non-identical tunnel diode
    elements. (a) fixed points for different values of $\epsilon$
    ($\epsilon_1=19.23$, $\epsilon_2=21.15$). (c) fixed points for
    different values of $a^{(2)}_1$ ($a^{(2)}=7.6570 V_0 -4.5468$,
    $a^{(2)}_2=7.6570 V_0 - 4.47032$). Again, solid lines are stable
    fixed points and dashed lines are unstable fixed points. The
    number along each branch marks the number of positive eigenvalues
    of the fixed point. The blue dots mark Hopf bifurcations, red
    squares are pitchfork bifurcations and green squares are fold
    bifurcations. (b) shows the break-up of bifurcations
    when changing the mismatch $\Delta\epsilon$ for different
    values of $\Delta\gamma$. Dashed line: Pitchfork bifurcation, red,
    green and blue line Hopf bifurcation for different values of
    $\Delta\gamma$. Parameters as in Fig.\ref{img:fix}.}
  \label{img:hettwo}
\end{figure}

In Fig.~\ref{img:hettwo}(a) the time scale parameter $\epsilon$ is
chosen as $\epsilon_1=19.23$ for the first subsystem and
$\epsilon_2=21.15$ for the second subsystem. \footnote{The value is
  calculated by introducing a difference of 10\% in the tunnel diode
  parameter $C_{d}$, which is the typical variance for general purpose
  tunnel diodes as denoted in the data sheet in \cite{GEN77}.}

The time-scale parameters do not influence the position of the fixed
points (see Eqs.~\eref{eq:modela}--~\eref{eq:modelc}) and thus the
curves in Fig.~\ref{img:hettwo}(a) and Fig.~\ref{img:fix} are
identical. Only the stability of the fixed points changes and the Hopf
bifurcation point shifts. This separates the Hopf and the pitchfork
bifurcation from each other.  A similar result can be obtained by
introducing heterogeneity in the parameter $\gamma$.

Fig.~\ref{img:hettwo}(b) shows this unfolding of the Hopf-pitchfork
bifurcation due to a difference $\Delta\epsilon=\epsilon_2-\epsilon_1$
in the parameter $\epsilon$. The dashed line corresponds to a
pitchfork bifurcation while the red, green and blue lines mark the
Hopf bifurcation curves in the $(V_0,\,\Delta\epsilon)$-plane. Each
color represents a different value of the difference in the timescale
parameter $\Delta\gamma=\gamma_2-\gamma_1$. In each case a
Hopf-pitchfork bifurcation can be obtained by using the correct set of
parameter values.

A mismatch in $a^{(2)}$, on the other hand, does influence the location
of the fixed points and in fact breaks up the pitchfork bifurcation 
and leads to an imperfect pitchfork bifurcation as discussed in
\cite{JUE97}. The pitchfork bifurcation separates into a stable branch
and a saddle-node bifurcation.  Figure~\ref{img:hettwo}(c) shows this
imperfect bifurcation. The parameter $a^{(2)}$ was previously set to
$a^{(2)}=7.6570 V_0 -4.5468$ for both elements. This value is used in
the first subsystem ($a^{(2)}_1$), while $a^{(2)}_2$ of the second
subsystem is set to $a^{(2)}_2=7.6570 V_0 - 4.47032$. \footnote{The
  value is calculated by introducing a difference of 1\% in the tunnel
  diode parameter $v_{d_0}$, which is the typical variance for high
  quality tunnel diodes.}

In the next sections we give an outlook into the dynamics of $N$
coupled tunnel diodes by performing a bifurcation analysis for a
system of four tunnel diodes in detail, and comparing the results to
electron transport models in superlattices.

\section{Arbitrary number of tunnel diodes in series}
\label{sec:arb}
In this section we perform a bifurcation analysis of four
identical tunnel diodes connected in series and aim to understand a
general serial array of $N$ tunnel diodes. Due to the high symmetry
the previously described Hopf-pitchfork bifurcation becomes a more
degenerate \cite{GOL85a} symmetry-breaking transition and we have to
distinguish between $N$ being even or odd.  These highly degenerate
bifurcations lead to multistability between different fixed point
branches and show close similarities to current branches in
superlattices. This similarity will be addressed below.

The linear stability analysis can be carried out in the same fashion
as for one and two tunnel diode elements. The general system
for the $z_n^*$ ($n=1,\ldots,N$) values of the fixed points ($x_1^*$,
$y_1^*$, $z_1^*$,$\ldots$,$x_N^*$, $y_N^*$,$z_N^*$) is given by
\begin{equation}
\label{eq:fixN}
 \frac{1}{3}(z_n^*)^3 -z_n^*- \frac{1}{d}\left(
   a^{(N)}-\sum^N_{m=1}z_m^* \right)=0, \qquad  n=1,\ldots,N,
\end{equation}
by using Eqs.~\eref{eq:modela}--\eref{eq:modelc}. System~(\ref{eq:fixN})
contains $N$ equations with $n=1,\ldots,N$. The Jacobian at the fixed
point is a block matrix
\begin{equation}
\label{eq:jacsN}
 \mathbf{J}^{(N)}=\left( \begin{array}{cccc}  \mathbf{J}^{(1)}_1 &  \mathbf{C} & \cdots & \mathbf{C}\\  \mathbf{C} &  \mathbf{J}^{(2)}_1 & &  \mathbf{C} \\ \vdots & & \ddots & \vdots \\  \mathbf{C} &  \mathbf{C} & \cdots &  \mathbf{J}^{(N)}_1  \end{array} \right),
\end{equation}
with
\begin{eqnarray}
 \mathbf{J}^{(i)}_1 &=& 
 \left( 
   \begin{array}{ccc}  
     1/\epsilon(1-(x_i^*)^2) & -1/\epsilon &0 \\  
     1 & 0 & 1\\  
     0& -d/\gamma&-1/\gamma 
   \end{array} 
 \right),  \label{eq:jacs1}\\
 \mathbf{C} &=& 
 \left( 
   \begin{array}{ccc}  
     0 & 0 & 0\\  
     0 & 0 & 0\\  
     0& 0&-1/\gamma 
   \end{array} 
 \right), \label{eq:couplingC}
\end{eqnarray}
where the blocks $\mathbf{C}$ account for coupling contributions, and
$x_n^*=-z_n^*$ ($n=1,\ldots,N$).

The system of $N$ tunnel diodes has an $S_N$-symmetry: the dynamical
equations are invariant under any permutation of the individual tunnel
diodes. I.e., if the indices of any two tunnel diodes are exchanged,
the equations are unchanged.  This is similar to the case of two
tunnel diodes in
Eqs.~\eref{eq:modela}--\eref{eq:modelc}. Additionally, we find the
previously discussed point symmetry around $a^{(N)} = 0$ in the
bifurcation diagram.

The analytic treatment of the Jacobian for $N$ diodes in
\ref{sec:appendixa} gives the following results. The symmetry-breaking
transitions occur at
\begin{equation*} 
  a^{(N)} = \pm (N-2d/3).
\end{equation*}
With increasing $a^{(N)}$ the symmetric fixed point becomes unstable at
$a^{(N)}=-(N-2d/3)$ and becomes stable again at $a^{(N)}=+(N-2d/3)$.  At
these transitions $N-1$ Hopf and $N-1$ zero-eigenvalue bifurcations
coincide. Thus $N-1$ real eigenvalues as well as $N-1$ complex
conjugate Hopf eigenvalue pairs simultaneously cross the imaginary
axis in the complex plane.

The $N$-dimensional system of equations for the fixed points
Eq.~\eref{eq:fixN} and the $3N$-dimensional Jacobian~\eref{eq:jacsN}
can be solved numerically for a given value of $N$. In the following,
we present the results for four tunnel diodes in series with the aim
to understand the experimental results in the next section. The
bifurcation pattern for an arbitrary number of tunnel diodes is then
explained based on these results.

\begin{figure}[ht]
 \includegraphics[width=\textwidth]{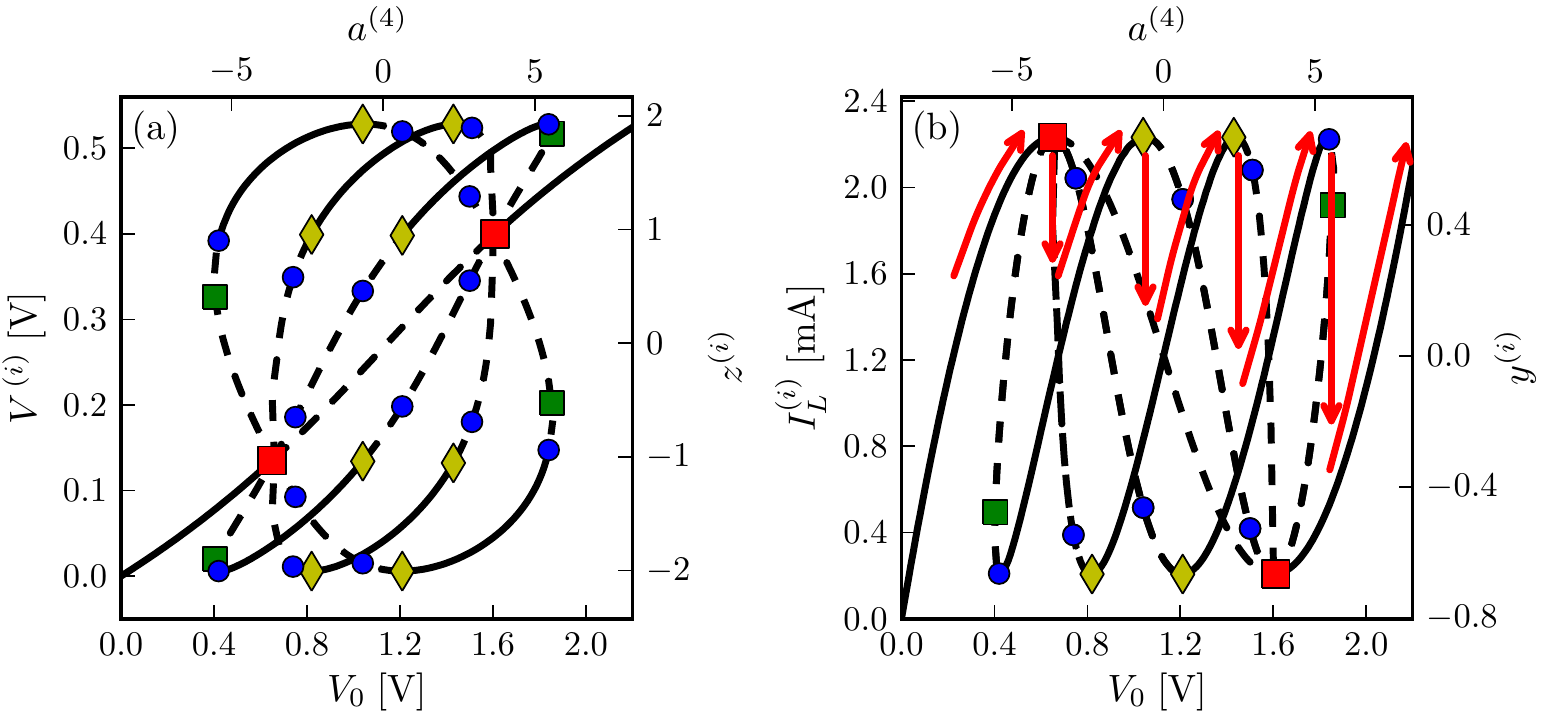}
 \caption{Voltage $V^{(i)}$ (a) and current $I_L^{(i)}$ (b) of fixed
   points according to Eqs.~(\eref{eq:modela}-\eref{eq:modelc}) for
   $N=4$ coupled tunnel diodes in dependence on the bias voltage $V_0$
   (only the voltages of primary branches are plotted). Black solid
   lines denote stable fixed points, while dashed lines are unstable
   fixed points. Red squares mark the symmetry-breaking transitions
   accompanied by Hopf bifurcations.  Blue dots, green squares, and
   yellow diamonds mark the location of Hopf bifurcations, fold
   bifurcations, and bifurcations of secondary branches accompanied by
   Hopf bifurcations, respectively.  Red arrows (b) illustrate the
   behavior of the current by applying a ramped voltage source
   (up-sweep). Parameters: $\epsilon=19.23$, $\gamma=23.18$, $d=0.57$,
   $a^{(4)}=7.66 V_0-8.64$.}
  \label{img:four}
\end{figure}
Figure~\ref{img:four}(a) shows a plot of the voltage $V^{(i)}$
variable of the fixed points for four tunnel diodes in series. This
bifurcation diagram holds for any of the four tunnel diodes
$i=1,\ldots,4$ due to the permutation symmetry. At $V_0 = 0.63
\mathrm{V}$ and at $V_0 = 1.62 \mathrm{V}$ (corresponding to the
$a^{(N)}$-values calculated in \ref{sec:appendixa}) the
symmetry-breaking transitions occur accompanied by Hopf bifurcations
(red squares).

The branches emanating at the transition point are generated as
follows.  Consider Eq.~\eref{eq:fixN} for the $z$-values of the fixed
points.  For certain fixed values of the sum $\sum_{m=1}^Nz_m$, the
cubic polynomial in this equation has three solutions. Thus the four
tunnel diodes can be distributed among these solutions, where the
exact values of $z_n$ and the sum follow from self-consistency.  This
type of cubic systems with $S_N$-symmetry have been studied from a
generic symmetry approach \cite{GOL02a} and the following results hold
for all such systems.

Apart from the symmetric branch, where all $z$-values are equal, there
are two other types of branches: Branches which have only two species,
i.e., each tunnel diode can be in either of two different states $z$,
are called \emph{primary branches}.  Solutions with three species, on
the other hand, are called \emph{secondary branches}. Primary branches
emanate from the symmetry-breaking point and are unstable for $N \ge
3$ close to the bifurcation point (even without the additional Hopf
bifurcations in our particular case). Primary branches can become
stable in two ways: They can bend back at a fold point (green squares
in Fig.~\ref{img:four}(a)) or they can be stabilized through
bifurcations with secondary branches (yellow diamonds in
Fig.~\ref{img:four}(a)). In our case due to the additional Hopf
bifurcations at the symmetry-breaking point, the primary branches have
to undergo another Hopf bifurcation in order to become stable
again. In the case of stabilization through bifurcations with
secondary branches, the additional Hopf bifurcations coincide with the
secondary bifurcations (yellow diamonds in Fig.~\ref{img:four}).  Note
that secondary bifurcation points can be found analytically (see
\ref{sec:appendixb}).

Let us now discuss in more detail the branch structure close to the
symmetry-breaking transition. In Fig.~\ref{img:four}(a) the possible
tunnel diode voltages are plotted. In this projection there are two
branches of transcritical type visible, i.e., branches which exist
below and above the bifurcation point.  Additionally there is
one branch of pitchfork type with a vertical tangent visible.

On the transcritical branches one tunnel diode is in one state (on the
leftmost branches for instance in the high voltage state) and three
are in the other state. How many of such solutions exist?  Since there
can be one of the four tunnel diodes in the single voltage state,
there are four such branches.  Note that in the projection of
Fig.~\ref{img:four} these branches coincide, since they have the
same tunnel diode voltages, only in a different order.  On the other
hand, on the pitchfork branch, two tunnel diodes are in the high
voltage and two tunnel diodes are in the low voltage state,
respectively. Since there are ${4 \choose 2}=6$ ways to choose two
from four tunnel diodes, there are 6 solutions of this pitchfork type,
which again coincide in the projection of Fig.~\ref{img:four}.

All of the above discussed solutions meet at the symmetry-breaking
bifurcation point.  Since a complete bifurcation analysis of all these
primary and secondary branches is beyond the scope of this work, we
focus on the experimentally observable features as shown in
Fig.~\ref{img:four}.

The stable branches have the shape of current branches in the current
component $I_L^{(i)}$ as shown in Fig.~\ref{img:four}(b). This means
that by increasing the applied voltage $V_0$ the current $I_L^{(i)}$
increases up to a threshold ($I_L^{(i)}=2.23 \mathrm{mA}$) where
$I_L^{(i)}$ suddenly drops to a lower current. This drop is related to
a jump of the voltage $V^{(i)}$ in one tunnel diode, i.e., from a low
voltage branch to a high voltage branch. After the jump the current
increases to the same threshold again. This behavior, which forms a
sawtooth each time, repeats $N$ times (red arrows in
Fig.~\ref{img:four}(b)). We note that similar current branches have
been reported for a very simple model consisting of a series array of
$NDC$ elements without any inductance \cite{XU10a}. The separate
voltage jump of each tunnel diode corresponds to the separate passing
through the NDC regime and as a result we observe current branches.

The symmetry-breaking transition is highly degenerate (see
\ref{sec:appendixa}) and gives rise to a plethora of branches. Due to
the coexistence of Hopf bifurcations each branch in
Fig.~\ref{img:four} is surrounded by limit cycles similar to the
system of two tunnel diode elements, as shown in Sec.~\ref{sec:two}.
When a current branch becomes unstable with increasing $V_0$ the
system may, instead of switching to the next branch, also switch to a
periodic orbit, due to the multistability between periodic orbits and
fixed points.

As discussed above for four tunnel diodes, we observe branches of
transcritical as well as of pitchfork type. From the investigation of
$N=5,\,6,\,7$, we found the following generic feature.  For odd $N$
all branches are of transcritical type, while for even $N$ we also
observe pitchfork type branches on which there are an equal number of
diodes in the upper and in the lower state. It is clear that for odd
$N$ the tunnel diodes cannot be divided into two equal numbers and
thus no pitchfork type branches are present.

With increasing $N$ the degeneracy of the bifurcation at the emanating
point increases due to the higher symmetry of the system and more
branches appear. For large $N$ the branch structure becomes very
similar to the current branches of superlattices. However, there is
one main difference between the two systems: In the tunnel diode
network there exists no determined order of switching of tunnel diodes
because of the previously discussed transformation invariance. This is
in contrast to the electronic transport models of superlattices, where
high- and low-field domains cause the current branches. High- and
low-field domains are groups of consecutive quantum wells, which are
either in a high-field state or in a low-field state. With increasing
voltage the high-field domain expands by one superlattice period,
i.e., the domain wall moves to the neighboring quantum well and forms
the next current branch, and so forth.  Thus the quantum wells switch
consecutively.

In the next section we compare our analytical and numerical results
for circuits containing a single, two, or four tunnel diodes to
measurements.

\section{Comparison of experimental and numerical results}
\label{sec:com}

In this section we compare the results from Secs.~\ref{sec:sin},
\ref{sec:two} and \ref{sec:arb} to experiments. As a setup for the
measurements, we use the circuit in Fig.~\ref{img:Ntds} with general
purpose devices including tunnel diodes of the type \texttt{1N3714},
ceramic capacitors with capacitance $C=56 \ \mathrm{pF} \pm 2 \
\mathrm{pF}$ and ceramic resistors with the resistance $R=47 \ \Omega
\pm 3 \ \Omega$. The parameters of the measurement circuit and the
parameters used in the simulations are listed in
Tab.~\ref{tab:comsin1}.

For the measurements with a ramped voltage source (increasing or
decreasing input voltage $V_0$), we use the arbitrary-function
generator Tektronix model \texttt{AFG3252}. We apply a slow frequency sawtooth-like
oscillation of 10~Hz such that we are able to observe stable
oscillating or fixed point states at each voltage. To decouple the
function generator from the circuit, we use a buffer, circuit based on a general purpose operational amplifier \texttt{LM741}.

We use the digital phosphor oscilloscope Tektronix model \texttt{DPO7104} in
combination with the differential probe \texttt{TD1000} (bandwidth 1 GHz) for measuring
the tunnel diode voltage or the active probe \texttt{TAP1500} (bandwidth 1 GHz) for
measuring the overall current.

In the simulations, a noise source is introduced to approximate noise
effects in the measurements. The noise is implemented into the
dynamical system by addition of an independent Gaussian white noise
term $D \xi^{(i)}(t)$ for each system with $D$ being the dimensionless noise intensity
set to $D=0.01$. The noise is added to the dynamical equation of the
$z_n$ variable of each subsystem, which is reasonable as $z_n$ is the rescaled voltage $V^{(n)}$.

\subsection{A single tunnel diode}

\begin{figure}[ht]
  \includegraphics[width=\textwidth]{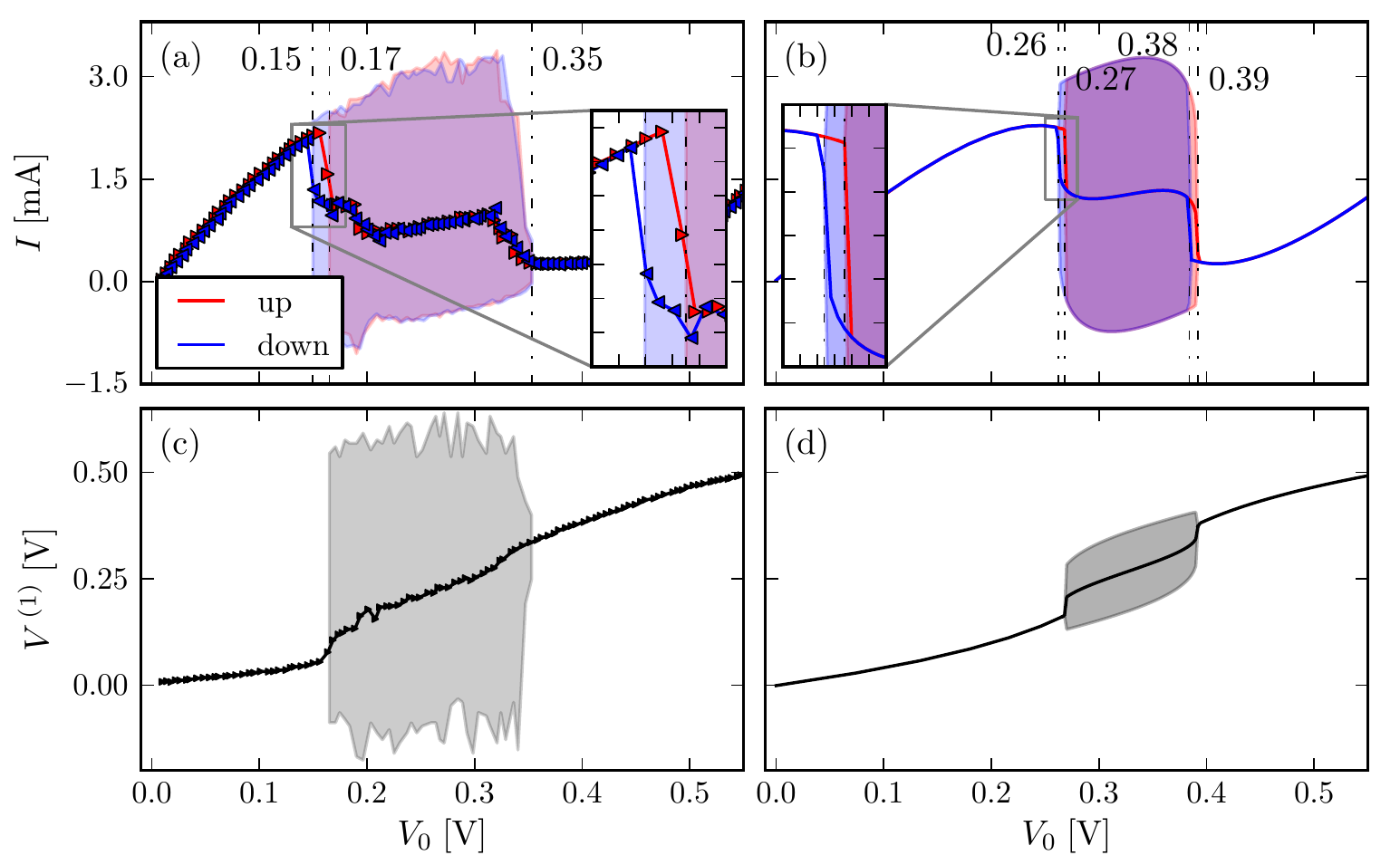}
 \caption{Single tunnel diode: Measured (a) and simulated (b) $I$-$V$ curve with 
applied voltage ramped upwards (red) and downwards (blue). 
The vertical dash-dotted lines (labeled by the voltages $V_0$)
indicate the boundaries of the oscillatory regimes for up- and down-sweep.
The insets show blow-ups of the hysteresis.
Lower Panel: Measured (c) and simulated (d) tunnel diode voltage $V^{(1)}$ during upwards ramping of $V_0$. Shaded areas show oscillatory regimes in all panels. Parameters used in the simulations are listed in Tab.~\ref{tab:comsin1}.}
  \label{img:vcom}
\end{figure}

The measured and simulated $I$-$V$ curves and voltages $V^{(1)}$ of a
single tunnel diode vs.~input voltage $V_0$ are shown in
Fig.~\ref{img:vcom}. The upper panels show the $I$-$V$ curve obtained
from measurements (panel (a)) and simulations (panel (b)). The lower
panels show the voltage $V^{(1)}$ from measurements (panel (c)) and
simulations (panel (d)). While the $I$-$V$ curve is plotted for upwards
and downwards ramping, the tunnel diode voltage is displayed for
upwards ramping only. The shaded area in all four panels shows the
range of time-dependent oscillations (range and amplitude of voltage $V_0$) 
with the lines denoting the time average of the oscillations in
this regime.

Movie 5 shows a plot similar to Fig.~\ref{img:vcom}(c), but ramping
the bias voltage both up- and downwards. This illustrates both the
oscillating regime (shaded in Fig.~\ref{img:vcom}) and the differences
between up- and downwards ramping caused by hysteresis. This
hysteresis is caused by the subcritical Hopf bifurcation as discussed
in Sec.~\ref{sec:sin}.

Overall, the simulated results are in good qualitative agreement with
the results from the experiments. The general characteristic, which
exhibits an increasing current $I$ or voltage $V^{(1)}$ followed by an
oscillatory regime and another increasing regime, is identical for
measurements and simulations. On the other hand, the oscillatory
regime and its amplitude differ. For example, during upwards ramping,
the measured oscillations start at $V_0 = 0.17$~V while in the
simulations oscillations occur after $V_0 = 0.27$~V, or the measured
voltage oscillations $V^{(1)}$ reach the negative voltage regime,
while the voltage oscillations stay positive in the simulations as
indicated in Fig.~\ref{img:vcom}. The different amplitudes of voltage
oscillations is due to the non-ideal nature of the experimental
voltage source. The differences in the bias voltage $V_0$ at which the
oscillating regime sets on can be explained by the restriction of the
third-order polynomial approximation of the $I$-$V$ curve in the
simulations. The parameters, as we fitted the curve, match the $I$-$V$
curve in the low- and high-$V_0$ regime, but as a trade-off the
location of the extrema of the curve -- which correspond to the onset
of oscillations -- is not perfectly reproduced by the fit as shown in
Fig.~\ref{img:bascufi}.

The typical $I$-$V$ characteristic of a single tunnel diode, which is
shown in Fig.~\ref{img:vcom}, starts with a single current branch and is followed
by a sudden collapse of the tunneling current because the energy
levels in a tunnel diode become misaligned. In the case of a load
resistor $R<\rho$ such that $0<d<1$, the collapse of the tunneling current corresponds to
oscillations. This means that when the tunnel diode passes through
the NDC regime it exhibits oscillations.

Simulations and measurements both show hysteresis while passing
through the NDC regime. The width of the hysteresis regime is $\Delta V_0
= 0.02 \ \mathrm{V} \pm 0.04 \ \mathrm{V}$ in experiments and in the simulations
$\Delta V_0 = 0.006 \ \mathrm{V}$. These results agree with the
previous results gained from the bifurcation analysis, where the
overlap between stable periodic orbits and stable fixed points has a
width of $\Delta V_0 = 0.006 \ \mathrm{V}$.  We attribute the difference in hysteresis ranges in measurement versus simulation to the use of a third order polynomial 
approximation in the simulations.

\subsection{Two tunnel diodes}
\begin{figure}[ht]
 \includegraphics[width=\textwidth]{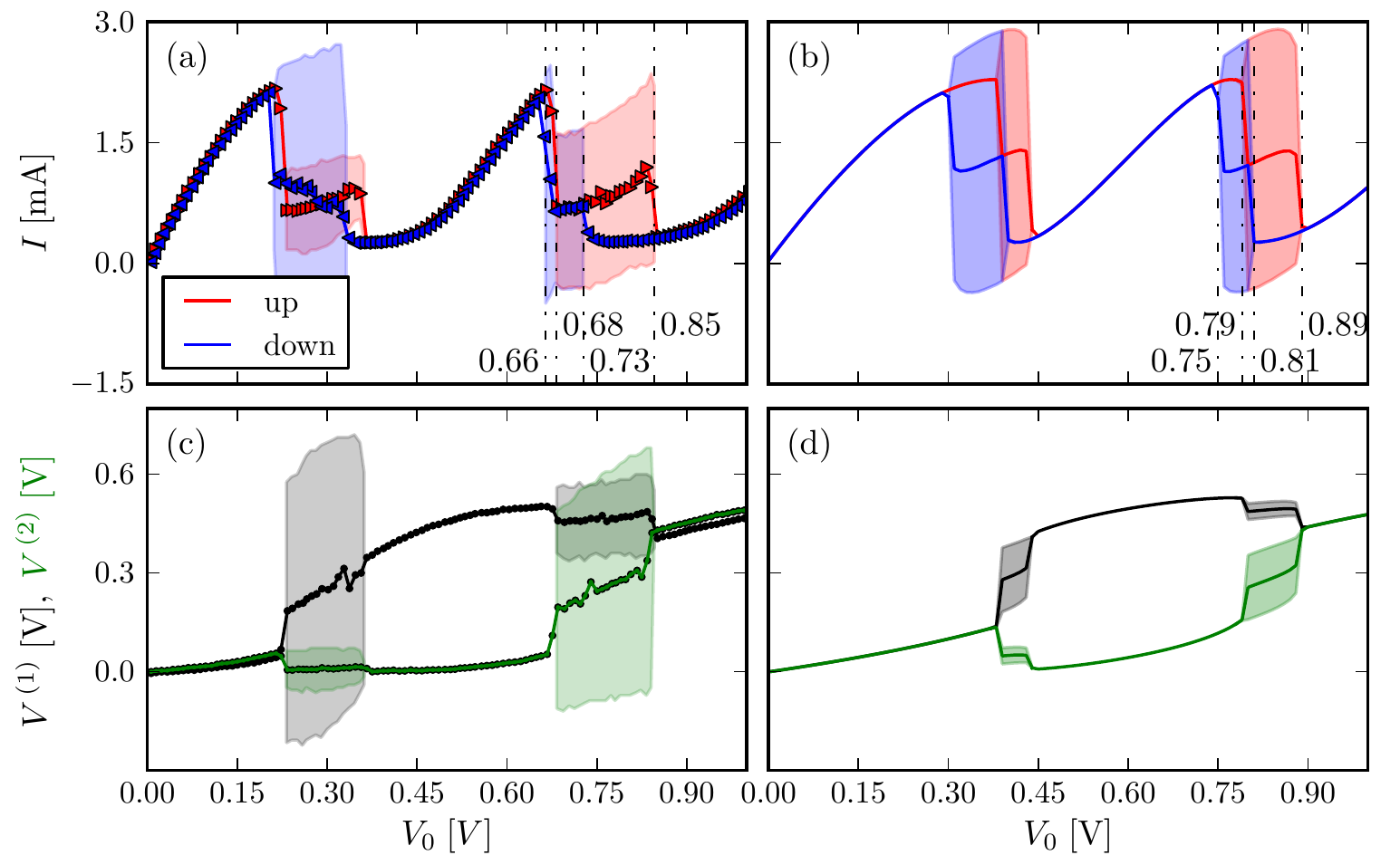}
 \caption{Two tunnel diodes in series: Measured (a) and simulated (b)
   $I$-$V$ curve with bias voltage $V_0$ ramped upwards (red) and
   downwards (blue). Lower panel: Measured (c) and simulated (d)
   tunnel diode voltages $V^{(1)}$ (black) and $V^{(2)}$ (green)
   during upwards ramping of $V_0$. Shaded areas show oscillatory regimes in all panels. Parameters used in the simulations are
   listed in Tab.~\ref{tab:comsin1}.}
  \label{img:v2com}
\end{figure}

The simulated and measured results of two tunnel diodes in series are
shown in Fig.~\ref{img:v2com}. Again the upper panels (a) and (b) show
the circuit $I$-$V$ curve during upwards (red) and downwards (blue)
ramping for measurements and simulations, respectively. The lower
panels are the tunnel diode voltages $V^{(1)}$ (black) and $V^{(2)}$
(green) during upwards ramping of the input voltage $V_0$ in
measurements (panel (c)) and simulations (panel (d)).

Movie 6 shows the effect of hysteresis in measurements on two coupled
tunnel diodes while ramping the bias voltage $V_0$ both up- and
downwards. The movie is otherwise similar to
Fig.~\ref{img:v2com}(c). The drift observable in this movie is not
caused by differences between the individual tunnel diodes
(cf. Fig.~\ref{img:hettwo}(c)), but influenced by the non-ideal
(i.e, not quasi-static) ramping in the measurements.

Again, the simulated and the measured results are qualitatively
similar. Simulation and measurement show hysteresis in the $I$-$V$
curve of both tunnel diodes, caused by the subcritical bifurcations,
which imply coexisting stable states. For example, during upwards
ramping of the applied voltage, the oscillations of the second
oscillation regime start in the measurement at $V_0 = 0.68 \
\mathrm{V}$ and ends at $V_0 = 0.85 \ \mathrm{V}$ while during
downwards ramping the oscillations occur between $V_0 = 0.66 \
\mathrm{V}$ and $V_0 = 0.73 \ \mathrm{V}$. In the simulations, the
oscillations of the second oscillation regime occur during upwards
ramping between $V_0 = 0.79 \ \mathrm{V}$ and $V_0 = 0.89 \
\mathrm{V}$ and during downwards ramping between $V_0 = 0.75 \
\mathrm{V}$ and $V_0 = 0.81 \ \mathrm{V}$ as indicated in
Fig.~\ref{img:v2com}.

Two current branches occur in the current-voltage characteristic of
two tunnel diodes in series (in panel (a) and (b) of
Fig.~\ref{img:v2com}). The tunneling current breaks down after
reaching a current threshold ($I_{measure} = 2.17 \ \mathrm{mA} \pm 0.02
\ \mathrm{mA} $, $I_{sim} = 2.28 \ \mathrm{mA}$) in both tunnel diodes
separately at different input voltages. This breakdown corresponds to
a pitchfork bifurcation in the bifurcation diagram. Prior to the
pitchfork bifurcation, both tunnel diodes are in the identical
tunneling state, while after the pitchfork bifurcation one tunnel
diode has passed through the NDC regime, while the other tunnel diode is still
in the tunneling state. These are the two asymmetric branches of the
pitchfork bifurcation. They can be seen in panel (c) and (d) of
Fig.~\ref{img:v2com}. The current branches occur simultaneously with
oscillations since the NDC regime is unstable, which gives rise to
oscillations. These oscillations are related to stable skew orbits as
discussed previously.

\subsection{Four tunnel diodes}
\begin{figure}[ht]
  \includegraphics[width=\textwidth]{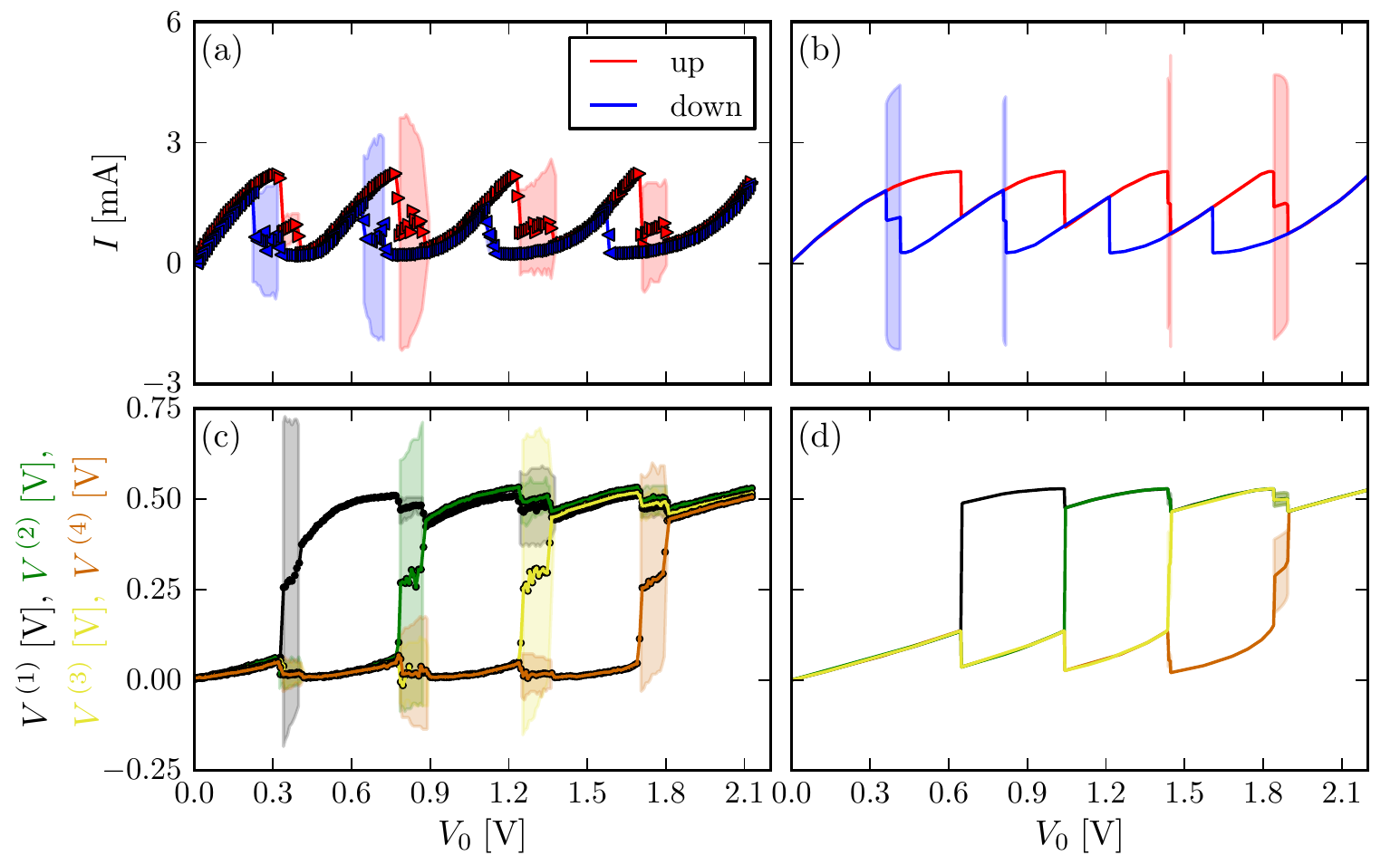}
  \caption{Four tunnel diodes in series: Measured (a) and simulated
    (b) $I$-$V$ curve with applied voltage ramped upwards (red) and
    downwards (blue). Lower panel: Measured (c) and simulated (d)
    tunnel diode voltages $V^{(1)}$ (black), $V^{(2)}$ (green),
    $V^{(3)}$ (yellow) and $V^{(4)}$ (brown) during upwards ramping of
    $V_0$. Shaded areas show oscillatory regimes in all
    panels. Parameters used in the simulations are listed in
    Tab.~\ref{tab:comsin1}.}
  \label{img:v4com}
\end{figure}

The simulated and measured results of four tunnel diodes in series,
which are shown in Fig.~\ref{img:v4com}, are similar to the results of
two tunnel diodes. Again the upper panels (a) and (b) show the circuit
$I$-$V$ curve during upwards (red) and downwards (blue) voltage
ramping for measurements and simulations, respectively. The lower
panels are the tunnel diode voltages $V^{(1)}$ (black), $V^{(2)}$
(green), $V^{(3)}$ (yellow) and $V^{(4)}$ (brown) during upwards
ramping of the input voltage $V_0$ in measurements (panel (c)) and
simulations (panel (d)).

In the case of four tunnel diodes in series, four current branches
occur as shown in Fig.~\ref{img:v4com}. Each tunnel diode switches
from a low-voltage state to a high-voltage state, separately and via
oscillations. Again, each tunnel diode passes through the NDC regime
at different levels of the bias voltage $V_0$. Both theoretical and
experimental investigations reproduce the described behavior as shown
in Fig.~\ref{img:v4com} also shown in movie 7, which resembles
Fig.~\ref{img:v4com}(c) but with up- and downwards ramping. Like in
movie 6, a drift is observable, which is caused by the non-ideal
(i.e., not quasi-static) ramping in the measurements.

The theoretical model describes the current branches as branches with
multiple degeneracy generated at symmetry-breaking transitions. Since
a tunnel diode can be at the upper or lower branch in
Fig.~\ref{img:four}(a), there exists no well-defined order of
switching of the tunnel diodes. The switching shown in
Fig.~\ref{img:v4com} is just one possible way determined by slight
differences between each tunnel diode. This is different from
electronic transport in superlattices, where the current branches are
caused by high- and low-field domains and consecutive shifting of the
domain wall.

\section{Conclusions}
\label{sec:con}
In this paper we have investigated a nonlinear circuit, which shows
dynamical scenarios similar to a semiconductor superlattice. As a
model for a sequence of tunneling barriers we use a series connection
of tunnel diodes, described by a three-variable extended
FitzHugh-Nagumo system. We have shown that a single tunnel diode
system reduces to the simplified FitzHugh-Nagumo system in the limit
case of small load resistance. Our numerical model predicts two
subcritical Hopf bifurcations in good agreement with measurements.

In the case of $N$ tunnel diodes in series, we have shown that the
model approximately describes electronic transport in superlattices
for certain parameter ranges and exhibits dynamics similar to that of
a superlattice. This includes multistable current branches, which
result in saw-tooth current-voltage characteristics and hysteresis
upon up- and down-sweep of applied voltage. As the cause for the
current branches we have found degenerate zero-eigenvalue
bifurcations.  Each of the branches corresponds to a different diode
passing through the NDC regime. The order in which the tunnel diodes
pass through the NDC regime is determined by small differences between
each of the devices. This is in contrast to semiconductor
superlattices, where the domain wall is shifted consecutively through
the superlattice.

Additionally, we have observed that each zero-eigenvalue bifurcation
occurs simultaneously with a Hopf bifurcation such that limit cycles
surround each fixed point branch. We have shown that this degenerate
bifurcation is followed by a pitchfork bifurcation of periodic orbits
in the case of two tunnel diode elements. We have been able to observe
this complex bifurcation scenario in measurements. The switching
between different branches of the pitchfork bifurcation is accompanied
by oscillations. The comparison with simulated results is in good
qualitative agreement.

The results of our work aim to understand and illustrate the complex
bifurcation scenario in a network with high symmetry, i.e., invariance
under any permutation of the tunnel diodes. We have shown that two
tunnel diodes which individually exhibit Hopf bifurcations can form a
Hopf-pitchfork bifurcation when coupled and that this degenerate
bifurcation is caused by the reflection symmetry of the system.  We
have also demonstrated that current branches can even occur in
networks without local coupling. A global coupling can already induce
zero-eigenvalue bifurcations, which are multiply degenerate and
results in multiple current branches. This is in contrast to
superlattices where local coupling is always present due to sequential
tunneling between neighboring quantum wells.

Our results may also help to understand the complex dynamics in cases 
where two or more tunnel diodes are used in series, for example in
multi-junction solar cells with more than two active solar subcells.

\appendix

\section{Analytic study of the bifurcations}

\subsection{Symmetry-breaking transition}
\label{sec:appendixa}

Consider the Jacobian (Eq.~\eref{eq:jacsN}) of the array of $N$ tunnel
diodes on a symmetric fixed point, i.e., all diodes are in the same
state.  In this case the Jacobian has the following structure
\begin{equation} 
  \mathbf{J}^{(N)} = \mathbf{J}_{1} \otimes \mathbf{I}_N + \mathbf{C} \otimes \mathbf{M}_N, \label{eq:tensor}
\end{equation}
where $\mathbf{J}_1$ is the Jacobian of a single diode, $\mathbf{C}$
is the coupling matrix, $\mathbf{I}_N$ is the $N$-dimensional identity matrix
and $\mathbf{M}_N$ is an $N\times N$ matrix given by
\begin{equation*} 
  \mathbf{M}_N =
  \left[
  \begin{array}{cccc}
    0      & 1      & \dots  & 1\\
    1      & \ddots & \ddots &\vdots\\
    \vdots & \ddots & \ddots & 1\\
    1      & \dots  &   1    & 0
  \end{array}
  \right].
\end{equation*}
The matrix $\mathbf{M}_N$ is a symmetric circulant matrix. It can be
diagonalized and has eigenvalues
\begin{equation*} 
  \mu_1 = N-1,\quad \mu_2=\mu_3=\dots=\mu_N=-1.
\end{equation*}
Note that the eigenvector corresponding to $\mu_1$ points in the
direction $(1,\,1,\dots,\, 1)$ and is thus tangent to the
synchronization manifold.  The eigenspace corresponding to the $(N-1)$
times degenerate eigenvalue $-1$, on the other hand, is perpendicular
to the synchronization manifold.

To find the eigenvalues $\lambda$ of $\mathbf{J}^{(N)}$ with the
characteristic equation
\begin{equation} 
  \det [\mathbf{J}^{(N)} - \lambda] = 0, \label{eq:char}
\end{equation}
we can diagonalize $\mathbf{M}_N$ in Eq.~\eref{eq:tensor} and since
the transformation does not affect the identity $\mathbf{I}_N$, the
characteristic equation Eq.~\eref{eq:char} is transformed into $N$
equations
\begin{equation*} 
  \det[\mathbf{J}_1 + \mu_k \mathbf{C} -\lambda] = 0 \qquad (k=1,\dots,N).
\end{equation*}
For $k=1$ this gives
\begin{equation} 
  \chi_{||}(\lambda):=\det[\mathbf{J}_1 + (N-1) \mathbf{C} -\lambda] = 0  \label{eq:chipar}
\end{equation}
and for all $(N-1)$ other $k$ we have
\begin{equation} 
  \chi_\perp(\lambda):=\det[\mathbf{J}_1 - \mathbf{C} -\lambda] = 0 . \label{eq:chiperp}
\end{equation}

A symmetry-breaking zero-eigenvalue transition then occurs if
\begin{equation} 
  \chi_\perp(0) = 0, \label{eq:symm-break-cond}
\end{equation}
and it is immediately clear that such a transition is $(N-1)$ fold
degenerate.  Using the explicit form of $\mathbf{J}_1$ and
$\mathbf{C}$ (see Eqs.~\eref{eq:jacs1} and \eref{eq:couplingC}) and
solving Eq.~\eref{eq:symm-break-cond} for $x$, we obtain
\begin{equation*} 
  x=\pm 1,
\end{equation*}
i.e., the symmetry-breaking transitions always occur at these values
of $x$ (see Fig.~\ref{img:autbo} and \ref{img:four}).  Inserting these
$x$-values into Eq.~\eref{eq:fixN} ($z_n=-x=\mp 1$) we can also find the
value of the bifurcation parameter $a^{(N)}$ at the symmetry-breaking
bifurcation:
\begin{equation} 
  a^{(N)} = \mp (N-2d/3)\qquad ({\rm for\; } x=\pm 1). \label{eq:bifparN}
\end{equation}

To find the overall eigenvalue structure at the symmetry breaking we
substitute $x=\pm 1$ into Eq.~\eref{eq:chiperp}. This gives
eigenvalues
\begin{equation} 
  \lambda = 0,\qquad \lambda = \pm \sqrt{-\frac{1}{\varepsilon}-\frac{d}{\gamma}} =  \pm i \sqrt{\frac{1}{\varepsilon}+\frac{d}{\gamma}} . \label{eq:ev-at-bif}
\end{equation}
Thus there is a Hopf bifurcation coinciding with the zero-eigenvalue
bifurcation and both of these are $(N-1)$ fold degenerate.

Note that it is possible to calculate the crossing direction of the
eigenvalues in the complex plane by implicit differentiation of
$\chi_\perp(\lambda)$ with respect to $a$. This shows that the fixed
point is transversely stable below the first transition
($a<-(N-2d/3)$) and above the second transition ($a>+(N-2d/3)$). Of
course this shows the stability only close to the transitions.

Let us now discuss the longitudinal stability of the fixed point. The
characteristic Eq.~(\ref{eq:chipar}) for the parallel direction
evaluated at $x=\pm 1$, is given by
\begin{equation*} 
 0 = b_3 \lambda^3 + b_2 \lambda^2 + b_1\lambda + b_0
\end{equation*}
with 
\begin{eqnarray*} 
b_3 &=& \varepsilon\gamma,\qquad b_2 = \varepsilon N,\qquad
b_1 = d\varepsilon + \gamma,\qquad b_0 = N.
\end{eqnarray*}
Since all coefficients are positive, the Routh-Hurwitz criterion, which
ensures that all eigenvalues are stable is given by
\begin{equation*} 
 0 < b_1 b_2 - b_0 b_3 = \varepsilon N(d\varepsilon + \gamma) - N\varepsilon\gamma =  d \varepsilon^2 N
\end{equation*}
and is always fulfilled ($d,N>0$). The synchronized fixed point is
thus stable in the parallel direction at the symmetry-breaking
transition and the bifurcation is observable for any number $N$ of
tunnel diodes.

\subsection{Bifurcations with secondary branches}
\label{sec:appendixb}

Consider Eq.~\eref{eq:fixN} for the $z$-coordinates of the tunnel diodes
\begin{equation} 
  f(z,\,\sigma):=\frac{1}{3}z^3-z-\frac{1}{d}(a-\sigma) = 0, \label{eq:zstart}
\end{equation}
where $\sigma = \sum_{m=1}^N z_m$. At the bifurcations between the
primary and the secondary branch, a change of two species (primary
branch) to three species (secondary branch) occurs. {\em Two species}
or {\em three species} in this context means that the tunnel diodes
are distributed among two or three different states,
respectively. Since this generation of a third species occurs
continuously, the bifurcation points are characterized by having a two
fold solution for $z$ in Eq.~\eref{eq:zstart}
\begin{eqnarray} 
 0 &=&  f(z,\, \sigma), \label{eq:f0}\\ 
 0 &=& \partial_z f(z,\,\sigma) \label{eq:Df0} .
\end{eqnarray}
Equation \eref{eq:Df0} gives as the location for the double root
\begin{equation*} 
  z = \pm 1.
\end{equation*}
Inserting this into Eq.~\eref{eq:f0} gives
\begin{equation} 
  \frac{1}{d}(a-\sigma) = \mp \frac{2}{3}, \label{eq:sigmaatbif}
\end{equation}
which can again be inserted into Eq.~\eref{eq:f0} to yield the
location of the other root
\begin{equation*} 
  z = \mp 2.
\end{equation*}

Let us now specifically investigate the case of four tunnel diodes
($N=4$).  To find the bifurcation parameter at the secondary
bifurcations, we need to determine the value of $\sigma$ at the
bifurcation.  In order for a secondary bifurcation to occur in the
first place, at least two tunnel diodes need to have $z=1$ or $z=-1$.
This gives the following possible values for $\sigma$
\begin{eqnarray} 
  \sigma &=& 2 \cdot (-1) + 2 \cdot (+2) = 2, \\
  \sigma &=& 2 \cdot (+1) + 2 \cdot (-2) = -2,\\
  \sigma &=& 3 \cdot (-1) + 1 \cdot (+2) = -1,\\
  \sigma &=& 3 \cdot (+1) + 1 \cdot (-2) = 1,
\end{eqnarray}
where the former two cases correspond to the branches of pitchfork
type and the latter two cases correspond to the branches of
transcritical type.  From Eq.~\eref{eq:sigmaatbif} we find 
possible values of $a$ at the bifurcation point
\begin{eqnarray} 
  a &=& +2 + 2d/3,\\
  a &=& +2 - 2d/3,\\
  a &=& -2 + 2d/3,\\
  a &=& -2 - 2d/3.
\end{eqnarray}
For other values of $N$ the secondary bifurcation points ($a$-values)
can be found similarly.

\subsection*{Acknowledgement}

We thank Pavel Rodin and John Neu for valuable discussions. This work
was supported in part by DFG in the framework of GRK 1558 and by the
U. S. National Science Foundation under Grant DMR-0804232.

\section*{References}


\end{document}


\subsection*{Supplementary material for the paper\\{\em Symmetry-breaking transitions in networks
of nonlinear circuit elements}}
{\large by M. Heinrich, T. Dahms, V. Flunkert, S. W. Teitsworth, and E. Sch{\"o}ll}\\

\textbf{Movie 1.}
\url{movie1.avi}

Hopf bifurcation of a single
tunnel diode. The bifurcation diagram (left) and the 3-dimensional
phase space (right) show fixed points and periodic orbits depending on
the input voltage $V_0$. The black solid and dashed lines (black dot
and white dot in phase space) correspond to stable and unstable fixed
points, respectively. The solid light-blue and dashed dark-blue lines
correspond to minimum and maximum values of stable and unstable
periodic orbits in the bifurcation diagram. The orbits are shown in
the phase space plot. Color shading in the bifurcation diagram shows
location and overlap of these periodic orbits. Parameters as given in
Tab.~2.\\

\textbf{Movie 2.}
\url{movie2.avi}

Bifurcations of two tunnel diodes
in series. The bifurcation diagram (left) and a 3-dimensional
projection of the 6-dimensional phase space (right) show fixed points
and periodic orbits depending on the input voltage $V_0$. The black
solid and dashed lines (black dot and white dot in phase space)
correspond to stable and unstable fixed points, respectively. Solid
and dashed green, orange and blue lines correspond to the maxima and
minima of periodic orbits, where solid and dashed lines denote stable
and unstable periodic orbits, respectively. The orbits are
shown in the phase space plot. Color shading in the bifurcation
diagram shows location and overlap of these periodic orbits. HB, HPF,
F, PPO are Hopf, Hopf-pitchfork, fold and pitchfork bifurcations of
periodic orbits, respectively. Parameters as given in Tab.~2.\\

\textbf{Movie 3.}
\url{movie3.avi}

Skew orbits of two tunnel
diodes in series in a 3-dimensional projection of the 6-dimensional
phase space. Elements of the Figure are as described in Movie 2. Both
tunnel diodes are oscillating at different orbits in
anti-phase. Parameters as given in Tab.~2, with $V_0=0.37$.\\

\textbf{Movie 4.}
\url{movie4.avi}

Anti-phase orbits of two
tunnel diodes in series in a 3-dimensional projection of the
6-dimensional phase space. Elements of the Figure are as described in
Movie 2. Both tunnel diodes are oscillating on the same orbit in
anti-phase. Parameters as given in Tab.~2, with $V_0=0.50$.\\

\textbf{Movie 5.}
\url{movie5.avi}

Measured tunnel diode voltage
$V^{(1)}$ of a single tunnel diode during upwards and downwards
ramping of $V_0$. A large amplitude signal shows oscillations. The
movie shows the hysteresis between upwards and downwards ramping.\\

\textbf{Movie 6.}
\url{movie6.avi}

Measured tunnel diode voltages
$V^{(1)}$ (black) and $V^{(2)}$ (green) of two tunnel diodes in series
during upwards and downwards ramping of $V_0$. A large amplitude
signal shows oscillations. The movie shows the hysteresis between
upwards and downwards ramping and the two asymmetric (skew-symmetric)
fixed points generated by pitchfork bifurcations. The switch from the
symmetric states (identical states of both tunnel diodes) to the
asymmetric state are accompanied by oscillations, which is explained
by the Hopf-pitchfork bifurcation.\\

\textbf{Movie 7.}
\url{movie7.avi}

Measured tunnel diode voltages $V^{(1)}$ (black), $V^{(2)}$ (green),
$V^{(3)}$ (yellow) and $V^{(4)}$ (brown) of four tunnel diodes in
series during upwards and downwards ramping of $V_0$. A large
amplitude signal shows oscillations. The movie shows the hysteresis
between upwards and downwards ramping and the symmetric as well as
asymmetric (skew-symmetric) fixed points generated by multiple
degenerate pitchfork bifurcations. Each tunnel diode is switching
between these states separately accompanied by oscillations. The order
of switching of the tunnel diodes shown here is just one out of many
possible switching orders.